\documentclass[journal,9pt]{IEEEtran}

\usepackage{graphicx}
\usepackage{authblk}
\usepackage{amsmath}
\usepackage{amssymb}
\usepackage{breqn}
\usepackage{scalerel}
\usepackage{relsize}
\usepackage{adjustbox}
\usepackage{xcolor}
\usepackage{mathtools}
\usepackage[compatibility=false]{caption}
\usepackage{subcaption}
\usepackage{silence}
\usepackage{float}
\usepackage{xcolor}
\colorlet{blue}{black} 
\usepackage{placeins} 
\usepackage{algorithm}
\usepackage{algorithmic}
\usepackage{lettrine}
\usepackage{url}
\newcommand{\e}[1]{\exp\!\left(#1\right)}
\newcommand{\norm}[1]{\left\lVert#1\right\rVert}
\newcommand{\abs}[1]{\left|#1\right|}
\newtheorem{lemma}{Lemma}
\newtheorem{theorem}{Theorem}
\newtheorem{remark}{Remark}
\usepackage{comment}

\begin{document}
\raggedbottom
\hbadness=10000
\vbadness=10000

\title{Two-Stage Rotation-Based Joint ToA-AoA Estimation for Spatial Wideband Systems}
\author{Chandrashekhar Rai and Debarati Sen 

\thanks{Both the authors are with G.S.Sanyal School of Telecommunications, Indian Institute of Technology (IIT) Kharagpur, India. Email: \protect\url{cs.rai93@iitkgp.ac.in}, \protect\url{debarati@gssst.iitkgp.ac.in}.}
\thanks{The conference precursor to this work has been published in the International Conference on Acoustic Speech and Signal Processing (ICASSP) 2025.}
}

\maketitle
\begin{abstract}
Joint delay-angle (time-of-arrival (ToA) and angle-of-arrival (AoA)) estimation in wideband multi-antenna orthogonal frequency division multiplexing (OFDM) systems is a key enabler for localization, sensing, and extremely large-multiple input multiple output (XL-MIMO) communications in mmWave/THz bands. In this regime, the spatial wideband effect (SWE), also known as range-angle coupling in MIMO frequency modulated continuous wave (FMCW) radar, breaks the conventional spatial narrowband assumption; the energy of each path is spread over multiple delay-angle bins, and the dominant peak of the 2D-DFT spectrum no longer coincides with the true coarse grid location. As a result, standard DFT-based coarse estimation followed by rotation/maximum-likelihood fine-tuning becomes inherently biased. In this paper, we first develop a rigorous finite-dimensional characterization of this phenomenon. Using a 2D Dirichlet-kernel model, we derive closed-form bounds on the SWE-induced spreading width and show that the significant energy of a single path is confined to at most $\lceil \alpha M |\tilde{\phi}_i| \rceil$ bins in both angle and delay, where $\alpha$ denotes the fractional bandwidth, $M$ the array size, and $\tilde{\phi}_i$ the normalized AoA of the $ i$-th scatterer. Building on this analysis, we propose a two-stage per-path rotation framework. In the first stage, we apply a cluster-level removal of the wideband coupling term and perform a localized 2D search to correct the SWE-induced coarse-bin shift. In the second stage, we show that the residual wideband term is uniformly bounded and can be treated as a small perturbation; this justifies a narrowband-type rotation refinement that yields super-resolution off-grid estimates of AoA, ToA, and path gains with near-DFT complexity. Extensive Monte-Carlo simulations on XL-MIMO OFDM models confirm that the proposed method achieves substantially higher hit rates, lower false-alarm rates, and lower RMSE than narrowband estimators, 2D-OMP, 2D-MUSIC, and direct wideband rotation schemes over a wide range of SNRs and fractional bandwidths, while maintaining scalability to large space-frequency grids.
\end{abstract}


\section{Introduction}
\textcolor{blue}{\lettrine{E}{stimation} of time of arrival (ToA) and angle of arrival (AoA) is critical for a wide range of emerging applications, including localization, sensing, wireless communications, and navigation in future 6G systems \cite{tataria20216g,tripathi2021millimeter,sarieddeen2021overview}. Joint ToA-AoA estimation is a fundamental enabler of several communication-centric functionalities, including beam management, beam alignment, initial access, mobility tracking, and channel prediction in beyond 5G (B5G) systems. In modern wideband MIMO-OFDM transceivers, accurate recovery of angular and temporal channel signatures directly impacts link establishment, spatial multiplexing, interference management, and low-overhead beam training. These parameters are also fundamental to autonomous driving, industrial safety, and smart home radar platforms \cite{kong2024survey}. The shift toward millimeter-wave (mmWave) and terahertz (THz) bands enables large-scale antenna arrays to be deployed in a compact aperture, thereby offering fine-grained angular resolution of multipath components in complex radio environments \cite{jornet2023wireless,ghasempour2017ieee}.} 

However, when orthogonal frequency-division multiplexing (OFDM) waveforms are combined with extremely large multiple-input multiple-output (XL-MIMO) arrays, the conventional finite-dimensional narrowband models become inaccurate. Specifically, the spatial wideband effect (SWE) causes frequency-dependent steering and induces beam-squint and delay-squint phenomena across the array \cite{jang2024learning}. A similar effect is observed in near-field communications where spherical wavefronts lead to range-dependent angle distortions \cite{wei2021channel,cui2022channel,lei2024deep}, which is out of scope for this work. In the radar literature, the same underlying coupling mechanism is referred to as the range-angle coupling (RAC) effect \cite{rabaste2013signal,durr2020range,hu2023range,han2023range,park2024spatial}. In addition, wideband DFT-based processing suffers from spectral leakage, which further degrades the reliability of 2D delay-angle peak detection. Therefore, to fully harness wideband multi-antenna and multicarrier systems for accurate ToA-AoA estimation, it is crucial to explicitly account for and mitigate these wideband-induced distortions.

\subsection{Related Works}

\subsubsection{ToA-AoA Estimation for MIMO-OFDM Systems}
A considerable volume of research exists on joint ToA-AoA estimation in wireless systems. Practical subspace-based localization techniques designed for commodity Wi-Fi platforms \cite{xiong2013arraytrack,kumar2014accurate,kotaru2015spotfi} demonstrate feasibility but are limited to small antenna dimensions due to hardware constraints. To better exploit sparsity in wideband channels, a compressive sensing (CS) framework is proposed in \cite{shahmansoori2017position}, where coarse signatures are extracted using simultaneous orthogonal matching pursuit (OMP) and refined using a high-complexity SAGE-based 3D-MUSIC spectrum search. With OFDM signaling, virtual array extension for accurate localization using only three antennas in a uniform circular array structure is introduced in \cite{chen2017joint}, while FFT-MUSIC and its two-step variant are proposed in \cite{chen2019low} to reduce computational burden.

Other works adopt separate 1D delay and angle recovery followed by pairing, enabling simplified near real-time processing \cite{ahmed2018estimating,pan2022efficient}. A fingerprinting-based location estimation for massive MIMO-OFDM is proposed in \cite{8307353}. Moreover, a rotation-based refinement is initially proposed for channel estimation in \cite{xie2016unified}, which further has been further extended to 3D AoA-aware sparse channel estimation in \cite{fan2018angle}. Our prior work \cite{rai2025low} leverages such rotation-assisted and 2D-OMP techniques to achieve low-complexity super-resolution ToA-AoA estimation, but under the spatial narrowband assumption.

Despite these advancements, all the aforementioned methods implicitly assume the conventional spatial narrowband assumption and do not account for the spatial wideband effect arising from ultra-wide bandwidths combined with massive/XL-MIMO arrays, leading to frequency-dependent array responses and coarse-bin misalignment. A principled treatment of SWE-aware modeling and low-complexity estimation under such regimes therefore remains an open and largely unexplored problem.

\subsubsection{Spatial Wideband Effect in MIMO-OFDM Systems}
The SWE has been identified as a critical challenge in massive MIMO-OFDM arrays \cite{wang2018spatial}, where beam-squint and delay-squint phenomena distort spatial-frequency processing and introduce spectral leakage for finite basis methods. Although \cite{wang2018spatial} proposes a leakage mitigation technique, its effectiveness still relies on accurate coarse delay-angle bin identification, which is not guaranteed under severe SWE. Moreover, \cite{wang2018spatial} does not discuss the clustering mechanism and the associated low signal-to-noise ratio (SNR) challenges. A machine learning-based cluster identification approach is introduced in \cite{rai2022signature} to improve robustness at low SNR, yet the cluster peak does not generally coincide with the true path index due to the spreading effect. Therefore, directly applying rotation-based techniques to mitigate spectral leakage effect as suggested in \cite{fan2018angle,wang2018spatial}, yields biased ToA-AoA estimates.

Moreover, optimization-based iterative joint delay-angle refinement methods \cite{wang2019beam} require high computational effort and are highly sensitive to initialization. Multidimensional subspace-based techniques \cite{weng2023wideband} additionally require subcarrier pairing via clustering, whose performance degrades substantially in low-SNR conditions. A sparse reconstruction under SWE is considered in \cite{rai2023sparse}, but delay and angle are treated separately, and paths do not share common ToA or AoA, limiting general applicability.

To the best of our knowledge, there is no existing work that rigorously quantifies the SWE-induced finite-bin spreading and coarse-bin shift, nor provides a provably correct coarse-to-fine AoA-ToA estimator with low complexity for massive/XL-MIMO wideband systems. This gap motivates the theoretical analysis and robust algorithm developed in this work.

\subsection{\textcolor{blue}{Contributions}}
\textcolor{blue}{The main contributions of this work are summarized below.}

\begin{itemize}
    \item \textcolor{blue}{\textbf{Finite-dimensional characterization of spatial wideband spreading and bin shift:} 
        We develop a new analytical framework that rigorously characterizes the effect of spatial wideband coupling on the 2D-DFT delay-angle domain. Using a Dirichlet-kernel representation, we derive closed-form bounds quantifying (i) the finite-bin spreading caused by fractional bandwidth and array size, and (ii) the resulting coarse-bin shift that biases all existing narrowband-based estimators. These results bridge the gap between continuous-time wideband propagation models and practical discrete-grid processing.}

            \textcolor{blue}{\item \textbf{SWE-aware coarse-bin realignment with cluster-level phase compensation:} 
                To correct the coarse-bin shift before refinement, we propose a novel localized coarse alignment strategy. It applies wideband phase compensation at the cluster level, followed by a restricted 2D search guided by the derived spreading bounds. This ensures correct grid localization with near-DFT complexity, enabling reliable coarse initialization even under strong SWE conditions.}

                    \item \textcolor{blue}{\textbf{Super-resolution wideband rotation refinement:} 
                        We show that after coarse realignment, the residual SWE term becomes uniformly bounded and can be treated as a small perturbation. Leveraging this fact, we extend narrowband rotation-based estimation to a wideband setting and derive closed-form refinement updates that yield super-resolution ToA-AoA estimates with significantly reduced computational cost compared to multidimensional subspace and CS-based methods.}

                            \item \textcolor{blue}{\textbf{Complexity and robustness advantages for XL-MIMO implementation:} 
                                A detailed complexity analysis demonstrates that the proposed algorithm scales linearly in the number of antennas and subcarriers, making it suitable for XL-MIMO hardware deployment. Furthermore, the framework avoids iterative global search or tensor decomposition, ensuring superior execution time, robustness to SNR variation, and compatibility with real-time processing constraints.}

                                    \item \textcolor{blue}{\textbf{Extensive validation on MIMO-OFDM models:} 
                                        Monte-Carlo simulations and numerical illustrations validate the theoretical claims, showing substantial improvements in hit rate, false-alarm rate, and RMSE over state-of-the-art approaches, including 2D-DFT, 2D-OMP, 2D-MUSIC, and direct wideband rotation schemes across a broad range of SNRs, array sizes, and fractional bandwidths. The proposed framework can be applied to both communication and sensing scenarios, such as mmWave/THz XL-MIMO-OFDM and automotive XL-MIMO-FMCW radar.}
                                        \end{itemize}

\subsection{Outline and Notations}
The subsequent sections of the paper are organized as follows: Section II presents the system model. A rotation-based ToA-AoA estimation technique for a spatial narrowband system is given in Section III as a preliminary. In Section IV, the suggested approach for estimating signatures in spatial wideband systems is introduced. Detailed numerical and simulation results are given in Section V, and Section VI concludes with the drawn insights. We use small unbold characters $a$ for scalar notions, small bold characters, $\textbf{a}$, for vector notions, and capital bold characters, $\textbf{A}$, for matrix notions. $ (.)^* $ denotes the conjugate, $(.)^T$ denotes the transpose, and $ (.)^H $ denotes the conjugate and transpose operation. $\lfloor x \rceil
$ denotes the closest integer number to a real number $x$.

\section{System Model} 

The baseband signal $x(t)$ scattered from I-point scatters\footnote{We assume point scatter for simplicity, and hence there will be a single physical path per scatter.}, received at the $m^{th}$ antenna of the receiver station (RS), can be written as

\par
\begin{figure}[h]
	\centering
	\includegraphics[width=0.98\linewidth]{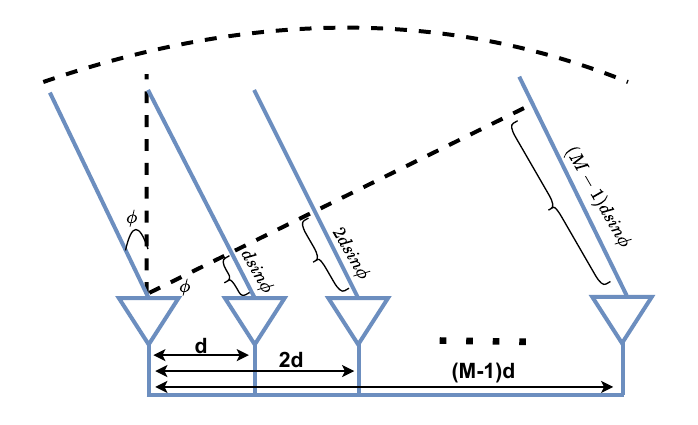}
	\caption{A ULA-based system model.}
	\label{fig_systemmodel}
\end{figure}

\noindent\small
\begin{align}
    y_m(t) = \sum\limits_{i = 0}^{I-1} \beta_i x(t-\tau_{m,i}) \e{-j2\pi f_c \tau_{m,i}}+w_m(t),
    \label{eq:discrete_physical_model}
\end{align}
\normalsize
where, $f_c$ is the carrier frequency, $\beta_i$ is the \textcolor{blue}{path gain coefficient (real-valued)} of the $i^{th}$ scatter,  $\tau_{m,i}$ is the delay of $i^{th}$ scatter at $m^{th}$ receive antenna which is given in \eqref{eq:delays}, for a specific ULA geometry shown in Fig.~\ref{fig_systemmodel}, and $w_m(t)$ is the additive white Gaussian noise (AWGN) with zero mean and $\sigma^2$ variance.

\par\noindent\small
\begin{align}
\color{blue}
\tau_{m,i}=\underbrace{\frac{r_{i}}{c}}_{part-I} + \underbrace{\frac{md.\sin\phi_{i}}{c}}_{part-II}.
\label{eq:delays}
\end{align}
\normalsize

In \eqref{eq:delays}, for an $i^{th}$ scatter -  delay in part-I ($\tau_{{\scaleto{0}{3pt}},i}$) is due to the radial distance $r_i$ between scatter and RS, delay in part-II ($\tau_{m_s,i}$) is due to the spatial delay induced from angle $\phi_i$ at the $m^{th}$ element of receiver antenna array with $d$ inter-element spacing. \textcolor{blue}{Since in this work, we are considering static scatters only, the delay due to mobility is ignored.}

From \eqref{eq:discrete_physical_model}, we can write the equivalent channel impulse response (CIR) of the measured wireless radio scene at the $m^{th}$ antenna as

\par\noindent\small
\begin{align}
    h_m(\tau) = \sum\limits_{i=0}^{I-1} \beta_i   \delta(\tau-\tau_{m,i})\e{-j2 \pi f_c \tau_{m,i}}+w_m(\tau).
    \label{eq:discrete_cir}
\end{align}
\normalsize

\subsubsection*{\textbf{Spatial Wideband System}} 
We can easily observe from \eqref{eq:delays}, operating at a wider BW along with a massive number of antenna sizes, restricts the system to non-avoidable spatial delay in the received signal along the aperture of ULA, i.e.,  $\tau_{0,i} \not\approx \tau_{M,i}$. For example, with $73$ GHz carrier frequency, $7$ GHz bandwidth, and $128$ antenna elements, the maximum spatial delay from a scatter located at $\phi = 60^{\circ}$ is approximately $5 T_s$ (here $T_s = 1/f_s$ is the \textcolor{blue}{sample duration}). This indicates that within one \textcolor{blue}{sample duration}, there are more than one different symbol across the aperture of ULA. 
Unlike conventional spatial narrowband systems, this situation demands special attention for the purpose of signature estimation that is further utilized in different applications like localization, beamforming, wireless communications, etc. \textcolor{blue}{We can rewrite the CIR in \eqref{eq:discrete_cir} using \eqref{eq:delays} as}

\par\noindent\small
\begin{align}
    h_m(\tau) = \sum\limits_{i=0}^{I-1} \beta_i \delta(\tau-\tau_{i}-\tau_{m_s,i})\e{-j2 \pi f_c \tau_{0,i}}\nonumber\\ \e{-j2 \pi f_c \tau_{m_s,i}}+w_m(\tau).
    \label{eq:discrete_physical_channel}
\end{align}
\normalsize

By defining $\Tilde{\beta}_i \triangleq
\beta_i e^{-j 2 \pi f_c \tau_{i}}$ and the normalized angle $\Tilde{\phi}_i \triangleq
dsin \phi_i /\lambda$, we can write \eqref{eq:discrete_physical_channel} as

\par\noindent\small
\begin{align}
    h_m(\tau) = \sum\limits_{i=0}^{I-1} \Tilde{\beta}_i   \delta(\tau-\tau_{{\scaleto{0}{3pt}},i}-\tau_{m_s,i}) \e{-j2 \pi m \Tilde{\phi}_i}+w_m(\tau).
    \label{eq:discrete_physical_channel1}
\end{align}
\normalsize

In order to observe the frequency response of the wireless channel, we take the continuous time Fourier transform (CTFT) of CIR in \eqref{eq:discrete_physical_channel1} w.r.t. $\tau$ as
\par\noindent\small
\begin{align}
    H_m(f) = \sum\limits_{i=0}^{I-1} \Tilde{\beta}_i \e{-j 2 \pi f \tau_{0,i}} \e{-j 2 \pi f \tau_{m_{s},i}}\nonumber\\ \e{-j 2 \pi m \Tilde{\phi}_i} + w_m(f).
\end{align}
\normalsize

We assume the system BW to be $f_s = \alpha f_c$, where $\alpha$ is the BW selection parameter.
Now, with the subcarrier spacing of $\Delta_f \triangleq f_s/N$, we can see the effect of OFDM signaling at the subcarrier level $\forall  n = 1 \cdots N$ and write the equivalent space-frequency channel response as
\par\noindent\small
\begin{align}
    H_m(n) = \sum\limits_{i=0}^{I-1} \Tilde{\beta}_i \e{-j 2 \pi n \Delta_f \tau_{0,i}} \e{-j 2 \pi m \Tilde{\phi}_i}\nonumber \\ \e{-j 2 \pi n \Delta_f m \frac{\Tilde{\phi}_i}{f_c}} +w_m(n),
    \label{eq:spatial_frequency_cir1}
\end{align}
\normalsize

where $w_m(n)$ are Independent and Identically Distributed (IID) noise samples that follow $\mathcal{CN}(0,\sigma^2)$.

We define normalized delay $\Tilde{\tau}_i \triangleq \Delta_{{\scaleto{f}{5pt}}} \tau_{{\scaleto{0}{3pt}},i}$ and write \eqref{eq:spatial_frequency_cir1} as 
\par\noindent\small
\begin{align}
    H_m(n) = \sum\limits_{i=0}^{I-1} \Tilde{\beta}_i &\e{-j 2 \pi n \Tilde{\tau}_i} \e{-j 2 \pi m \Tilde{\phi}_i} \e{-j 2 \pi n \Delta_f  m \frac{\Tilde{\phi}_i}{f_c}}\nonumber\\ &+w_m(n).
    \label{eq:spatial_frequency_cir}
\end{align}
\normalsize

Since \textcolor{blue}{$\Delta_f mn\frac{\Tilde{\phi}_i}{f_c} = \frac{f_s}{N} mn \frac{\Tilde{\phi}_i}{f_c} = \frac{\alpha f_c}{N} mn \frac{\Tilde{\phi}_i}{f_c} = \frac{\alpha}{N} mn \Tilde{\phi}_i$} , we can write the discrete space-frequency spatial wideband channel as
\par\noindent\small
\begin{align}
    H_m(n) = \sum\limits_{i=0}^{I-1} \Tilde{\beta}_i &\e{-j 2 \pi n \Tilde{\tau}_i} \e{-j 2 \pi m \Tilde{\phi}_i} \underbrace{\e{-j 2 \pi \frac{\alpha}{N} mn\Tilde{\phi}_i}}_{Wideband-Term}\nonumber \\&+w_m(n).
    \label{eq:discrete_space_frequency_channel}
\end{align}
\normalsize

From \eqref{eq:discrete_space_frequency_channel}, we observe that for $\alpha \rightarrow 0 $, the wideband-term vanishes to unity, and the conventional temporal-only wideband system can be shown as a special case of the spatial-temporal wideband system.
Finally, we define $\lbrace
\Tilde{\tau}_i,\Tilde{\phi}_i, \Tilde{\beta}_i, I \rbrace $ as the normalized signature of the wireless radio scene that is to be estimated.

\begin{remark}
    \label{remark_radar}
    \textit{To provide a comprehensive understanding, Appendix~\ref{app_radar_model} presents a detailed derivation of a radar signal model that explicitly incorporates the RAC effect for XL-MIMO-frequency modulated continuous wave (FMCW) radar systems, which is mathematically equivalent to the SWE in XL-MIMO-OFDM systems (except the negative sign in the phase). This extended formulation highlights the inherent coupling between range and angular parameters in scenarios involving extremely large antenna arrays and wide operational bandwidths. Furthermore, the proposed estimation framework developed in this work is inherently adaptable and can be seamlessly applied to achieve unbiased range–angle parameter estimation in  XL-MIMO-FMCW radar configurations that simultaneously leverage large-scale array apertures and high-bandwidth operation.}
\end{remark}

\section{\textcolor{blue}{Preliminaries}}
\label{sec_pre}

\textcolor{blue}{In this section, we introduce the rotation-based framework for off-grid joint ToA-AoA estimation under the spatial narrowband assumption. The goal is to establish the coarse-to-fine estimation procedure that later serves as the foundation for the spatial wideband extension in Section~\ref{sec_wideband}. For $\alpha \rightarrow 0$, the discrete space–frequency channel model reduces to}
\par\noindent\small
\begin{align}
H(m,n)=\sum_{i=0}^{I-1}\Tilde{\beta}_i\, 
\e{-j2\pi m\Tilde{\phi}_i}\,
\e{-j2\pi n\Tilde{\tau}_i}
+w(m,n),
\label{eq:narrowband_physical_channel}
\end{align}
\normalsize
\textcolor{blue}{which reveals that joint ToA-AoA estimation essentially becomes a joint two-dimensional frequency estimation problem.}

\subsection{\textcolor{blue}{2D-DFT Representation and Grid Mismatch}}

\textcolor{blue}{DFT-based spectrum estimation is particularly attractive in XL/massive MIMO settings due to its low computational cost and suitability for hardware implementation. Moreover, when $M$ and $N$ are large, DFT grids provide sufficiently fine resolution, whereas subspace- or CS-based approaches become computationally prohibitive. To develop a principled coarse-to-fine estimation framework, we first characterize how a finite-dimensional $M \times N$ space-frequency response maps into the discrete delay-angle domain via a 2D-IDFT operation. In particular, we seek an explicit expression for the resulting 2D spectrum magnitude in terms of the underlying continuous (off-grid) signatures.}

\begin{lemma}
\label{lemma_1}
\textcolor{blue}{For finite $(M,N)$, the magnitude of the 2D angle–delay response is}
\end{lemma}
\par\noindent\small
\begin{align}
\textcolor{blue}{
\abs{G(k,l)} 
= 
\abs{
\sum_{i=0}^{I-1}
\frac{\Tilde{\beta}_i}{\sqrt{MN}}
D_M\!\left(\Tilde{\phi}_i-\frac{k}{M}\right)
D_N\!\left(\Tilde{\tau}_i-\frac{l}{N}\right)
}.
}
\end{align}
\normalsize
\textcolor{blue}{where $\abs{D_Q(x)}\doteq\abs{\sin(Q\pi x)/\sin(\pi x)}$, and asymptotically 
$\lim_{Q\to\infty }\big(\!\frac{1}{\sqrt{Q}}\abs{D_Q(x)}\big)
= \sqrt{Q}\,\delta(x)$.}

\vspace{2mm}
\begin{proof}
    \textcolor{blue}{Follows directly by taking the 2D-IDFT of \eqref{eq:narrowband_physical_channel}.}
\end{proof}

\vspace{2mm}
\textcolor{blue}{In the asymptotic regime $(M,N\rightarrow\infty)$, Lemma~\ref{lemma_1} yields}
\par\noindent\small
\begin{align}
\textcolor{blue}{\abs{G(k,l)} = 
\sqrt{MN}
\sum_{i=0}^{I-1}\Tilde{\beta}_i\,
\delta\!\left(\Tilde{\phi}_i-\frac{k}{M}\right)
\delta\!\left(\Tilde{\tau}_i-\frac{l}{N}\right),
\label{eq_asymptotic}
}
\end{align}
\normalsize
\textcolor{blue}{so the $i$th scatter ideally lies at bin $(k,l)$ for 
$\Tilde{\phi}_i=k/M$, $\Tilde{\tau}_i=l/N$.  
In practice, however, $M$ and $N$ are finite and grid mismatch is unavoidable: unless the signature lies exactly on a DFT bin, the corresponding energy leaks to neighboring bins according to 
$D_M(\cdot)$ and $D_N(\cdot)$.  
Thus, every non-integer (NI) signature can be written as}
\par\noindent\small
\begin{align}
\textcolor{blue}{
\Tilde{\phi}_i^{\text{NI}} 
= \Tilde{\phi}_i^{\text{I}} + \Tilde{\phi}_i^{\text{F}},\qquad
\Tilde{\tau}_i^{\text{NI}} 
= \Tilde{\tau}_i^{\text{I}} + \Tilde{\tau}_i^{\text{F}}.
}
\end{align}
\normalsize
\textcolor{blue}{Since most of the energy remains concentrated around 
$(\lfloor M\Tilde{\phi}_i\rceil,\lfloor N\Tilde{\tau}_i\rceil)$, 
the peak of $\abs{G(k,l)}$ provides a coarse estimator of the grid-aligned signature.  
(Later, we show that this intuition breaks under the spatial-wideband effect.)}

\subsection{\textcolor{blue}{Scatter Detection and Coarse Signature Estimation}}

\textcolor{blue}{The number of scatter-induced clusters in the delay–angle domain can be identified using clustering-based approaches \cite{rai2025artificial}, support-detection techniques \cite{bevelander2025divide}, or dedicated peak-search methods \cite{rai2025low}.  
Assuming that clusters do not overlap under the SWE condition, each scatter generates a distinct peak in the 2D spectrum.  
We employ the 2D peak-search algorithm of \cite{MATLABCode} to detect the dominant peaks of $\abs{G(k,l)}$, yielding an estimate of the number of scatters $\hat{I}$.  
The peak coordinates give the coarse signature for the $i$th scatter:}
\par\noindent\small
\begin{align}
\textcolor{blue}{
\widehat{\Tilde{\phi}_i^{I}} = \frac{\hat{k}_i}{M},\qquad
\widehat{\Tilde{\tau}_i^{I}} = \frac{\hat{l}_i}{N}.
}
\end{align}
\normalsize


\subsection{\textcolor{blue}{Why One-Bin Separation is Realistic in mmWave/THz massive MIMO Systems}}

\textcolor{blue}{
The proposed framework is motivated by mmWave/THz massive MIMO systems, where the measurement grid is naturally large (typical configurations use $(M,N)\geq128$). For a $128\times128$ space–frequency grid, the resulting angular resolution is  
$1/M \approx 0.0078$ (i.e., $\sim0.45^\circ$), and the temporal resolution $1/N\approx7.8\times10^{-3}$ corresponds to sub-nanosecond delays under GHz-level bandwidths, giving centimeter-level range resolution at $50$–$100$\,m.
}

\textcolor{blue}{
In such propagation environments, two resolvable scatters typically differ by one or more DFT bins in at least one domain.\footnote{\textcolor{blue}{Scatterers with angular separation $<0.4^\circ$ or delay separation of only a few tens of picoseconds would be inseparable at these frequencies.}}  
For arrays with fewer antennas or narrower bandwidths, parametric super-resolution (MUSIC/ESPRIT/CS) becomes necessary.  
However, applying such algorithms to a massive $128\times128$ grid is computationally prohibitive for real-time communication use-cases.  
Therefore, the DFT-based coarse–fine structure adopted here is both practical and well aligned with mmWave/THz system requirements.
}

\textcolor{blue}{
Under the spatial narrowband assumption, the coarse peak locations  
$(\hat{k}_i,\hat{l}_i)$ directly correspond to the integer parts of the signature.  
The fractional offsets can be recovered via a rotation-based refinement procedure, which we develop next.  
This narrowband rotation mechanism forms the basis for the two-stage wideband estimator presented in Section~\ref{sec_wideband}.
}

\subsection{\textcolor{blue}{Rotation-Based Fine-Tuned Estimation (Narrowband Case)}}

\textcolor{blue}{
After obtaining the coarse grid-aligned signatures $(\hat{k}_i,\hat{l}_i)$ for each detected scatter, the remaining task is to refine the fractional offsets 
$(\Tilde{\phi}_i^{F},\Tilde{\tau}_i^{F})$.
To achieve this, we employ a per-path rotation mechanism that aligns the $i$th scatter with its nearest 2D-DFT atom in the local neighborhood of its coarse bin.}

\textcolor{blue}{
For any trial offsets $(\Delta\Tilde{\phi},\Delta\Tilde{\tau})$, the rotated channel for the $i$th path is defined as}
\vspace{0.5mm}
\par\noindent\small
\begin{align}
\textcolor{blue}{
H_i^{\mathrm{rot}}(m,n;\Delta\Tilde{\phi},\Delta\Tilde{\tau})
=
\e{j2\pi m\,\Delta\Tilde{\tau}}\;
H(m,n)\;
\e{j2\pi n\,\Delta\Tilde{\phi}}.
}
\label{eq:rot_op_integrated}
\end{align}
\normalsize

\textcolor{blue}{
Let $\mathbf{f}_M(k)\!\in\!\mathbb{C}^{M}$ and $\mathbf{f}_N(l)\!\in\!\mathbb{C}^{N}$ denote the $M$- and $N$-point unit-norm DFT steering vectors}
\par\noindent\small
\begin{align*}
\textcolor{blue}{
\mathbf{f}_M(k)=\frac{1}{\sqrt{M}}
\left[1,\,\e{j2\pi\frac{k}{M}},\dots,\e{j2\pi\frac{(M-1)k}{M}}\right]^T,
}
\end{align*}
\begin{align*}
\textcolor{blue}{
\mathbf{f}_N(l)=\frac{1}{\sqrt{N}}
\left[1,\,\e{j2\pi\frac{l}{N}},\dots,\e{j2\pi\frac{(N-1)l}{N}}\right]^T.
}
\end{align*}
\normalsize

\textcolor{blue}{
Evaluating the rotated channel at the coarse bin gives the refinement objective}
\vspace{0.5mm}
\par\noindent\small
\begin{align}
\textcolor{blue}{
J_i(\Delta\Tilde{\phi},\Delta\Tilde{\tau})
=
\left|
\mathbf{f}_M^{H}(\hat{k}_i)\,
H_i^{\mathrm{rot}}(\cdot;\Delta\Tilde{\phi},\Delta\Tilde{\tau})\,
\mathbf{f}_N(\hat{l}_i)
\right|.
}
\label{eq:rot_objective_integrated}
\end{align}
\normalsize

\textcolor{blue}{
Under the narrowband model, the contribution of the $i$th path to $J_i$ takes the form}
\par\noindent\small
\begin{align}
\textcolor{blue}{
\begin{aligned}
J_i(\Delta\Tilde{\phi},\Delta\Tilde{\tau})
&=
\frac{|\Tilde{\beta}_i|}{\sqrt{MN}}\,
\Bigg|
D_M\!\left(\Tilde{\phi}_i-\Delta\Tilde{\tau}-\frac{\hat{k}_i}{M}\right)
\\[-2pt]
&\qquad\qquad\qquad\quad\times~
D_N\!\left(\Tilde{\tau}_i-\Delta\Tilde{\phi}-\frac{\hat{l}_i}{N}\right)
\Bigg|.
\end{aligned}
}
\label{eq:local_singlepath_integrated}
\end{align}
\normalsize

\textcolor{blue}{
Each Dirichlet kernel in \eqref{eq:local_singlepath_integrated} attains its peak when the trial offsets match the true fractional offsets, i.e., when both kernel arguments are zero. 
Thus, fine estimation reduces to the local ML optimization}
\vspace{0.5mm}

\par\noindent\small
\begin{equation}
\textcolor{blue}{
(\widehat{\Delta\Tilde{\phi}}_i,\widehat{\Delta\Tilde{\tau}}_i)
=
\arg\max_{\Delta\Tilde{\phi},\,\Delta\Tilde{\tau}}
~~J_i(\Delta\Tilde{\phi},\Delta\Tilde{\tau}),
}
\label{eq:ml_rotation_integrated}
\end{equation}
\normalsize
\textcolor{blue}{
where the search is restricted to}
\vspace{0.5mm}

\par\noindent\small
\[
\textcolor{blue}{
\Delta\Tilde{\phi}\in\!\left[-\tfrac{1}{2N},\tfrac{1}{2N}\right],~~
\Delta\Tilde{\tau}\in\!\left[-\tfrac{1}{2M},\tfrac{1}{2M}\right].
}
\]
\normalsize

\textcolor{blue}{
A hierarchical binary refinement strategy, as proposed in the conference version of this work \cite{rai2025low}, efficiently finds the low-complexity maximizer of \eqref{eq:ml_rotation_integrated}.}

\vspace{2mm}

\begin{theorem}[{\textcolor{blue}{Optimality of the Local Refinement}}]
\label{thm:rotation_optimal_integrated}
\textcolor{blue}{
Let $(\hat{k}_i,\hat{l}_i)$ be the coarse bin of the $i$th scatter.
Under the narrowband model, the refinement objective $J_i(\Delta\Tilde{\phi},\Delta\Tilde{\tau})$ defined in \eqref{eq:rot_objective_integrated} satisfies:}
\begin{enumerate}
\item \textcolor{blue}{
The $i$th path contribution} 
\vspace{0.5mm}
\par\noindent\small
\begin{equation}
\textcolor{blue}{
\begin{aligned}
J_i(\Delta\Tilde{\phi},\Delta\Tilde{\tau})
&=
\frac{|\Tilde{\beta}_i|}{\sqrt{MN}}\,
\Bigg|
D_M\!\left(\Tilde{\phi}_i-\Delta\Tilde{\tau}-\frac{\hat{k}_i}{M}\right)
\\[-4pt]
&\hspace{37pt}\times~
D_N\!\left(\Tilde{\tau}_i-\Delta\Tilde{\phi}-\frac{\hat{l}_i}{N}\right)
\Bigg|
\end{aligned}
}
\end{equation}
\normalsize

\textcolor{blue}{achieves a unique maximum at the true fractional offsets 
$(\Tilde{\phi}_i^{F},\Tilde{\tau}_i^{F})$.}

\item \textcolor{blue}{
For any scatter $j\neq i$, the rotated contribution inside the $i$th refinement region satisfies}
\vspace{0.5mm}
\par\noindent\small
\begin{align}
\textcolor{blue}{
\frac{
\left|G_j^{\mathrm{rot}}(\hat{k}_i,\hat{l}_i;\Delta)\right|
}{
\left|G_i^{\mathrm{rot}}(\hat{k}_i,\hat{l}_i;\Delta)\right|
}
=\mathcal{O}\!\left(\frac{1}{M}\right)
~~\text{or}~~
\mathcal{O}\!\left(\frac{1}{N}\right),
}
\end{align}
\normalsize
\textcolor{blue}{since the desired path aligns with the Dirichlet main-lobe while all other paths contribute only bounded sidelobes.}

\item \textcolor{blue}{
As $(M,N)\to\infty$, the refinement objective converges to}
\par\noindent\small
\begin{align}
\textcolor{blue}{J_i(\Delta\Tilde{\phi},\Delta\Tilde{\tau})
\;\xrightarrow{}\;
|\Tilde{\beta}_i|\sqrt{MN}\,
\delta\!\Big(\Delta\Tilde{\tau}-\Tilde{\tau}_i^{F}\Big)\,
\delta\!\Big(\Delta\Tilde{\phi}-\Tilde{\phi}_i^{F}\Big),
}
\end{align}
\normalsize
\textcolor{blue}{recovering the exact signature in the asymptotic limit.}
\end{enumerate}
\end{theorem}

\begin{proof}
\textcolor{blue}{
Substituting the rotated channel \eqref{eq:rot_op_integrated} into 
$J_i(\Delta)$ yields the decomposition}
\par\noindent\small
\begin{align}
\textcolor{blue}{J_i(\Delta)=
\left|
G_i^{\mathrm{rot}}(\hat{k}_i,\hat{l}_i;\Delta)
+
\sum_{j\neq i}
G_j^{\mathrm{rot}}(\hat{k}_i,\hat{l}_i;\Delta)
\right|.
}
\end{align}
\normalsize
\textcolor{blue}{Evaluating the $i$th term leads to \eqref{eq:local_singlepath_integrated}, whose maximum occurs only when both Dirichlet arguments vanish, establishing part~1.  
For part~2, because $(\hat{k}_i,\hat{l}_i)\neq(k_j,l_j)$ for $j\neq i$, 
each $j$th contribution is restricted to Dirichlet sidelobes of bounded magnitude, while the desired term attains a main-lobe magnitude of order $\mathcal{O}(M)$ and $\mathcal{O}(N)$, proving the stated ratio bound.  
Part~3 follows from Lemma~\ref{lemma_1}, using the well-known limit 
$\frac{1}{\sqrt{Q}}|D_Q(x)|\rightarrow\sqrt{Q}\,\delta(x)$ as $Q\to\infty$.}
\end{proof}

\section{Proposed Methodolgy}
\label{sec_wideband}

Unlike the spatial narrowband case discussed in Section~\ref{sec_pre}, the spatial SWE fundamentally alters the structure of the 2D delay-angle
spectrum. In particular, the energy of a single propagation path is no longer concentrated around a single dominant delay-angle bin. Instead, due to the
wideband coupling term, each path produces a finite two-dimensional spread (cluster) in the delay-angle domain, and the dominant peak inside this
cluster may shift away from the nearest DFT grid point corresponding to the true coarse signature. Therefore, to enable a reliable coarse-to-fine estimation framework under SWE, we must first identify a localized search region around the detected peak (or cluster) of each path and then realign the coarse delay-angle bin within this region. This requires a finite-dimensional characterization of the SWE-induced spreading width in both delay and angle, which is given in the following theorem. 

\begin{theorem}[Spreading Limits Under SWE]
\label{thm:SWE_spreading}
\textcolor{blue}{%
Consider the single-path spatial wideband model in~\eqref{eq:discrete_space_frequency_channel} with normalized AoA 
$\widetilde{\phi}_i$ and normalized ToA $\widetilde{\tau}_i$. 
Let $G_i(k,l)$ denote the $M\times N$ angle-delay response of this path. 
For any fixed threshold $0<\varepsilon<1$, define the set of significant delay bins at the
coarse angle bin $\hat{k}_i$ as}
\begin{equation}
\color{blue}
\mathcal{L}_i(\varepsilon)
\triangleq
\Big\{ l\in\{0,\dots,N-1\} :
\frac{|G_i(\hat{k}_i,l)|}{\max_{l'}|G_i(\hat{k}_i,l')|}
\ge \varepsilon \Big\},
\label{eq:Li_def_SWE}
\end{equation}
\textcolor{blue}{%
and define $\mathcal{K}_i(\varepsilon)$ analogously in the angle domain. 
Then there exists a constant $C(\varepsilon)$, independent of $M$ and $N$, such that}
\begin{equation}
\color{blue}
|\mathcal{K}_i(\varepsilon)| \le \left\lceil \alpha M|\widetilde{\phi}_i|\right\rceil + C(\varepsilon),
\qquad
|\mathcal{L}_i(\varepsilon)| \le \left\lceil \alpha M|\widetilde{\phi}_i|\right\rceil + C(\varepsilon),
\label{eq:Li_Ki_bounds_SWE}
\end{equation}
\textcolor{blue}{%
i.e., the spatial wideband induced spreading in both angle and delay is upper bounded, up to an
$O(1)$ edge term, by the closed-form quantity}
\begin{equation}
\color{blue}
L_i \triangleq \left\lceil \alpha M|\widetilde{\phi}_i|\right\rceil.
\label{eq:Li_closed_form_SWE}
\end{equation}
\textcolor{blue}{%
Moreover, as $M,N\to\infty$ with fixed fractional bandwidth $\alpha$, we have
$|\mathcal{K}_i(\varepsilon)|,|\mathcal{L}_i(\varepsilon)| \sim \alpha M|\widetilde{\phi}_i|$,
which characterizes the spreading bins due to SWE.}
\end{theorem}

\begin{proof}
    \textcolor{blue}{Given in Appendix~\ref{thm:SWE_spreading}}.
\end{proof}

\begin{remark}
    \label{remark_1}
  \textit{ \textcolor{blue}{Theorem~\ref{thm:SWE_spreading} shows that, under the spatial wideband effect, the contribution of a single path is no longer concentrated at a single angle–delay bin, but is spread over a block of approximately $L_i=\lceil \alpha M|\widetilde{\phi}_i|\rceil$ bins in each domain. This spreading is governed by the Dirichlet kernels $D_M(\cdot)$ and $D_N(\cdot)$ (cf.\ Lemma~\ref{lemma_1}), whose main lobes and sidelobes determine how the path energy is distributed within the block. As a consequence, the largest samples of $|G(k,l)|$ inside a highly spread block do \emph{not} necessarily occur at the DFT grid point closest to the true normalized signature $(\widetilde{\phi}_i,\widetilde{\tau}_i)$, i.e., at $(k,l) = (\lfloor M\widetilde{\phi}_i\rceil,\lfloor N\widetilde{\tau}_i\rceil)$. In other words, under strong SWE the dominant peak in a block is generally a result of the Dirichlet shaping and coupling, and cannot be interpreted as a direct coarse estimate of the true DoA–ToA.}}
\end{remark} 

\begin{figure}[htbp]
    \centering
    \includegraphics[width=0.7\linewidth]{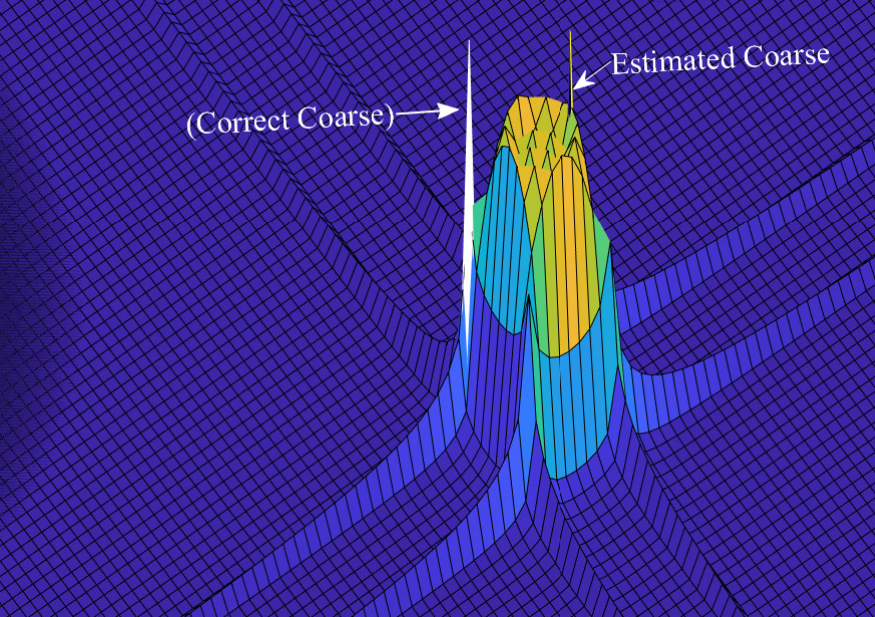}
    \caption{Spatial wideband effect and coarse-bin shifting at $\alpha = 0.15$.}
    \label{fig:bin_shift}
\end{figure}

\textcolor{blue}{%
To illustrate the practical implication of Theorem~\ref{thm:SWE_spreading}, consider  a single path with true normalized signature  $(\widetilde{\phi}_i,\widetilde{\tau}_i)=({80.25}/{M},{88.50}/{N})$ in a $128\times128$ MIMO-OFDM system. Under the narrowband assumption, the dominant peak would ideally appear at the nearest DFT grid point $(k,l)=(80,89)$. However, when $\alpha=0.15$, the spatial wideband coupling 
causes the largest magnitude sample of the 2D DFT spectrum to shift to $(82,95)$ as shown in Fig.~\ref{fig:bin_shift}. This example confirms that, when the spreading width $L_i\simeq \alpha M\widetilde{\phi}_i$ becomes non-negligible, the energy is redistributed within a block, and the maximum bin inside the block ceases to correspond to the true coarse signature. In other words, relying solely on the peak location leads to an inherently biased 
coarse DoA-ToA estimate under SWE.}
\textcolor{blue}{
Table~\ref{tab_bin_shift} further quantifies this effect for different fractional bandwidths $\alpha$. For $\alpha\leq0.01$, the spreading is small and the estimated coarse bins remain aligned with the true grid location. In contrast, for $\alpha\geq0.05$, noticeable bin shifts occur even in the noise-free case, demonstrating that coarse localization must operate at the cluster level rather than the peak level.
}

\begin{table}[htbp]
\caption{Coarse-Bin shifting in the spatial wideband systems for a scatter with $(\Tilde{\phi}_i,\Tilde{\tau}_i) = (80.25/M,88.50/N)$.}
\label{tab_bin_shift}
\resizebox{\columnwidth}{!}{%
\begin{tabular}{|c|c|c|c|c|}
\hline
                             & \textbf{$\alpha = 0.01$} & \textbf{$\alpha = 0.05$} & \textbf{$\alpha = 0.1$} & \textbf{$\alpha = 0.2$} \\ \hline
\textbf{Estimated Angle-Bin} & 80                       & 83                       & 88                      & 81                      \\ \hline
\textbf{Estimated Delay-Bin} & 89                       & 91                       & 93                      & 96                      \\ \hline
\end{tabular}%
}
\end{table}

\textcolor{blue}{%
Once a cluster associated with the $i$th path is identified, we suppress the induced spatial wideband distortion using a conjugation operation based on the initial coarse AoA estimate $\widehat{\widetilde{\phi}}_i=\hat{k}_i/M$,}
\par\noindent\small
\begin{align}
\textcolor{blue}{
H_i^{\rm rem}(m,n)
=
\widetilde{\beta}_i
\e{-j2\pi m\widetilde{\phi}_i}
\e{-j2\pi n\widetilde{\tau}_i}}\nonumber \\
\textcolor{blue}{\e{-j2\pi \frac{\alpha}{N}mn(\widetilde{\phi}_i-\widehat{\widetilde{\phi}}_i)}},
\label{eq:Hrem}
\end{align}
\normalsize
\textcolor{blue}{%
which ideally eliminates the SWE distortion when the coarse angle is accurate. Any remaining misalignment is then refined through the rotation-based fine estimation developed in Section~\ref{sec_pre}. Thus, two issues must be addressed sequentially:
(i) correcting the coarse-bin shift induced by SWE, and (ii) mitigating residual leakage from the fractional components of the signature.}

\textcolor{blue}{%
Motivated by the above observations, we propose a novel two-stage rotation-based scheme for accurate signature estimation under spatial wideband conditions. In the first stage, we perform a coarse rotation centered at the dominant peak of each identified cluster to recover the correct coarse angle–delay bin that has shifted due to the spatial wideband effect. In the second stage, a refined rotation is applied around this corrected coarse bin to compensate for the residual fractional offsets, thereby achieving high-precision off-grid ToA-AoA estimation. As demonstrated later in Section~\ref{sec_results}, this two-stage approach significantly outperforms the state-of-the-art direct-rotation method of \cite{wang2018spatial} for spatial wideband systems, particularly in regimes where the spreading width $L_i=\lceil \alpha M|\widetilde{\phi}_i|\rceil$ becomes non-negligible.}

\subsection{\textcolor{blue}{First-Stage Rotation-Coarse Bin Correction}}

\textcolor{blue}{%
Due to spatial wideband induced spreading (Theorem~\ref{thm:SWE_spreading}), the dominant peak inside a cluster does not generally correspond to the true coarse signature. Therefore, the first stage aims to realign the cluster and recover the correct coarse angle–delay bin for each detected path. Let $(\hat{k}_i^{(0)},\hat{l}_i^{(0)})$ be the initial peak detected within the $i$th path cluster $\mathcal{C}_i$ in the 2D spectrum $G(k,l)$.
We define the initial coarse DoA estimate as $\widehat{\widetilde{\phi}}_i^{(0)} = \hat{k}_i^{(0)}/M.$}

\textcolor{blue}{%
Using this coarse estimate, we construct the conjugation factor that counters the SWE:}
\par\noindent\small
\begin{align}
\textcolor{blue}{
C_i(m,n)
=
\exp\!\left(+j2\pi\frac{\alpha}{N}\,mn\,\widehat{\widetilde{\phi}}_i^{(0)}\right).}
\label{eq:Ci}
\end{align}
\normalsize

\textcolor{blue}{%
We then rotate the entire space-frequency channel by}
\vspace{1mm}
\par\noindent\small
\begin{align}
\textcolor{blue}{
H^{(1)}(m,n)
=
H(m,n)\,C_i(m,n),}
\label{eq:H1_def}
\end{align}
\normalsize
\textcolor{blue}{%
which eliminates most of the coupling term 
$e^{-j2\pi(\alpha/N)mn\widetilde{\phi}_i}$ for the $i$th path. After the wideband conjugation in \eqref{eq:H1_def}, the cluster maximum may still be displaced from the true coarse grid location due to SWE-induced Dirichlet reshaping. To recover the correct coarse bin, we perform a localized discrete rotation over the cluster neighborhood. Let $\mathcal{C}_i$ denote the detected cluster of the $i$th path, and define a bounded search region}
\par\noindent\small
\begin{align}
\color{blue}
\mathcal{S}_i \triangleq \{(k,l): |k-\hat{k}_i^{(0)}|\le L_i,\; |l-\hat{l}_i^{(0)}|\le L_i\},
\end{align}
\normalsize
\textcolor{blue}{where $L_i=\lceil \alpha M|\tilde{\phi}_i|\rceil$ is the SWE-induced spreading width (Theorem~\ref{thm:SWE_spreading}).  
For each candidate $(k,l)\in\mathcal{S}_i$, we apply a bin-centered rotation}
\par\noindent\small
\begin{align}
\color{blue}
H^{(1)}_{i,(k,l)}(m,n)
=
\e{j2\pi m\frac{k}{M}} H^{(1)}(m,n)\,
\e{j2\pi n\frac{l}{N}},
\end{align}
\normalsize
\textcolor{blue}{
and evaluate the DC component of its 2D-IDFT. The corrected coarse bin is then obtained as}
\par\noindent\small
\begin{align}
\color{blue}
(\hat{k}_i^{(c)},\hat{l}_i^{(c)})
=
\arg\max_{(k,l)\in\mathcal{S}_i}
\left|
\mathbf{f}_M^{H}(k)\,H^{(1)}_{i,(k,l)}\,\mathbf{f}_N(l)
\right|.
\end{align}

\textcolor{blue}{The updated coarse signature is thus}
\par\noindent\small
\begin{align}
\textcolor{blue}{
\widehat{\widetilde{\phi}}_i^{(c)} = \frac{\hat{k}_i^{(c)}}{M},\qquad
\widehat{\widetilde{\tau}}_i^{(c)} = \frac{\hat{l}_i^{(c)}}{N}.}
\label{eq:corrected_coarse}
\end{align}
\normalsize

\textcolor{blue}{%
These corrected coarse estimates serve as the initialization for the second-stage rotation,
which refines the fractional offsets and recovers the high-precision ToA-AoA estimates.}

\subsection{\textcolor{blue}{Second-Stage Rotation-Fine Off-Grid Estimation}}

\textcolor{blue}{%
After the first-stage compensation, each path cluster is re-centered so that the dominant
energy is aligned with the corrected coarse bin $(\hat{k}_i^{(c)},\hat{l}_i^{(c)})$.
However, a residual spatial wideband term remains due to the mismatch between the true DoA
$\widetilde{\phi}_i$ and the corrected coarse estimate
$\widehat{\widetilde{\phi}}_i^{(c)} = \hat{k}_i^{(c)}/M$.
In this subsection, we explicitly characterize this residual term, justify when it can be
neglected, and then apply the narrowband rotation principle from Section~\ref{sec_pre}
to obtain fine off-grid ToA-AoA estimates.}

\vspace{2mm}
\noindent\textbf{\textcolor{blue}{Residual spatial wideband term:}}

\textcolor{blue}{%
Using the corrected coarse angle $\widehat{\widetilde{\phi}}_i^{(c)}$, we define the
compensated channel as}
\par\noindent\small
\begin{equation}
\color{blue}
H_i^{(c)}(m,n)
=
H(m,n)\,
\exp\!\left(+j2\pi\frac{\alpha}{N}mn\,\widehat{\widetilde{\phi}}_i^{(c)}\right),
\label{eq:Hc_def}
\end{equation}
\normalsize
\textcolor{blue}{%
where $H(m,n)$ is the full space-frequency channel in~(9). Substituting~(9) into
\eqref{eq:Hc_def} gives}
\par\noindent\small
\begin{align}
\color{blue}
H_i^{(c)}(m,n)
=
\sum_{i=0}^{I-1}
\widetilde{\beta}_i\,
&\color{blue}\e{-j2\pi m\widetilde{\phi}_i}
\e{-j2\pi n\widetilde{\tau}_i}\nonumber\\
&\color{blue} \e{-j2\pi\frac{\alpha}{N}mn(\widetilde{\phi}_i-\widehat{\widetilde{\phi}}_i^{(c)})}
+ w^{(c)}(m,n).
\label{eq:Hc_expand}
\end{align}
\normalsize

\textcolor{blue}{%
For the $i$th path, the corresponding signal term is}
\vspace{1mm}
\par\noindent\small
\begin{align}
\color{blue}
H_{i,{\rm sig}}^{(c)}(m,n)
=
\widetilde{\beta}_i\,
\e{-j2\pi m\widetilde{\phi}_i}
\e{-j2\pi n\widetilde{\tau}_i}
\nonumber\\ \textcolor{blue}{\e{-j2\pi\frac{\alpha}{N}mn\Delta\widetilde{\phi}_i}},
\label{eq:Hc_sig}
\end{align}
\normalsize
\textcolor{blue}{%
where $\Delta\widetilde{\phi}_i \triangleq \widetilde{\phi}_i-\widehat{\widetilde{\phi}}_i^{(c)}$
denotes the residual off-grid error after coarse correction. The multiplicative factor}
\par\noindent\small
\begin{align}
\color{blue}
R_i(m,n)
\doteq
\exp\!\left(-j2\pi\frac{\alpha}{N}mn\Delta\widetilde{\phi}_i\right)
\label{eq:Ri_def}
\end{align}
\normalsize
\textcolor{blue}{%
represents the residual spatial wideband distortion.}

\vspace{2 mm}
\noindent\textbf{\textcolor{blue}{Neglecting the residual term:}}

\textcolor{blue}{%
From the coarse quantization, we have $|\Delta\widetilde{\phi}_i|\le 1/(2M)$.
For $(m,n)\in\{0,\ldots,M-1\}\times\{0,\ldots,N-1\}$, the residual phase satisfies}
\par\noindent\small
\begin{align}
\color{blue}
\left|
2\pi\frac{\alpha}{N}mn\Delta\widetilde{\phi}_i
\right|
\le
2\pi\alpha\,
\frac{(M-1)(N-1)}{N}\,
\frac{1}{2M}
\approx
\pi\alpha,
\label{eq:residual_phase_bound}
\end{align}
\normalsize
\textcolor{blue}{%
so that for moderate fractional bandwidths (e.g., $\alpha\le 0.1$), the residual term
$R_i(m,n)$ induces at most a small phase rotation. Using the first-order expansion
$\e{-j\theta}\approx 1-j\theta$ for $|\theta|\ll1$, we obtain}
\par\noindent\small
\begin{align}
\color{blue}
R_i(m,n)
\approx 1 - j\,2\pi\frac{\alpha}{N}mn\Delta\widetilde{\phi}_i,
\label{eq:Ri_approx}
\end{align}
\normalsize
\textcolor{blue}{%
which shows that the dominant behavior of $H_{i,{\rm sig}}^{(c)}(m,n)$ is governed by the
narrowband term $\widetilde{\beta}_i \e{-j2\pi m\widetilde{\phi}_i}\e{-j2\pi n\widetilde{\tau}_i}$, while the residual part acts as a small perturbation. In the local neighborhood of the corrected coarse bin $(\hat{k}_i^{(c)},\hat{l}_i^{(c)})$ where the fine-tuning is performed, this perturbation has negligible impact on the location of the maximum of the rotation objective. Hence, we approximate}
\par\noindent\small
\begin{align}
\color{blue}
H_{i,{\rm sig}}^{(c)}(m,n)
\approx
\widetilde{\beta}_i\,
\e{-j2\pi m\widetilde{\phi}_i}
\e{-j2\pi n\widetilde{\tau}_i},
\label{eq:Hc_narrowband_equiv}
\end{align}
\normalsize
\textcolor{blue}{%
which corresponds to an equivalent spatial narrowband model for the $i$th path.}

\begin{algorithm}
\footnotesize
\caption{Two-Stage Rotation-Based ToA-AoA and Coefficient Estimation under SWE}
\label{alg:2stage_swe}
\begin{algorithmic}[1]
\REQUIRE Space-frequency channel $\mathbf{H}\in\mathbb{C}^{M\times N}$, fractional bandwidth $\alpha$.
\ENSURE $\{\widehat{\widetilde{\phi}}_i,\widehat{\widetilde{\tau}}_i,\widehat{\widetilde{\beta}}_i\}_{i=1}^{\hat{I}}$.

\STATE \textbf{2D IDFT and cluster detection}
\STATE Compute $\mathbf{G}=\mathrm{IDFT}_M(\mathrm{IDFT}_N(\mathbf{H}))$.
\STATE Detect $\hat{I}$ clusters $\{\mathcal{C}_i\}_{i=1}^{\hat{I}}$ in $|\mathbf{G}|$ and, for each $\mathcal{C}_i$, find initial peak
$(\hat{k}_i^{(0)},\hat{l}_i^{(0)})$.
\STATE Set $\widehat{\widetilde{\phi}}_i^{(0)}=\hat{k}_i^{(0)}/M$.

\STATE \textbf{First-stage rotation (coarse bin correction)}
\FOR{$i=1$ to $\hat{I}$}
  \STATE Form the wideband conjugation factor
  \[
  C_i(m,n)=\exp\!\Big(j2\pi\frac{\alpha}{N}mn\,\widehat{\widetilde{\phi}}_i^{(0)}\Big).
  \]
  \STATE Compute the conjugated channel
  \[
  H_i^{(1)}(m,n)=H(m,n)\,C_i(m,n).
  \]
  \STATE Define the SWE-bounded search region
  \[
  \mathcal{S}_i=\{(k,l):|k-\hat{k}_i^{(0)}|\le L_i,\;|l-\hat{l}_i^{(0)}|\le L_i\},
  \]
  where $L_i=\lceil \alpha M|\widehat{\widetilde{\phi}}_i^{(0)}|\rceil$.
  \STATE \textbf{Discrete coarse rotation across bins}
  \FORALL{$(k,l)\in\mathcal{S}_i$}
    \STATE Form the bin-rotated channel
    \[
    H_{i,(k,l)}^{(1)}(m,n)
    =
    H_i^{(1)}(m,n)\,
    e^{j2\pi\left(m\frac{k}{M}+n\frac{l}{N}\right)}.
    \]
    \STATE Compute
    \[
    J_i^{(c)}(k,l)
    =
    \Big|
    \mathbf{f}_M^{H}(k)\,
    H_{i,(k,l)}^{(1)}\,
    \mathbf{f}_N(l)
    \Big|.
    \]
  \ENDFOR
  \STATE Obtain the corrected coarse bin
  \[
  (\hat{k}_i^{(c)},\hat{l}_i^{(c)})
  =
  \arg\max_{(k,l)\in\mathcal{S}_i}
  J_i^{(c)}(k,l).
  \]
  \STATE Set
  $\widehat{\widetilde{\phi}}_i^{(c)}=\hat{k}_i^{(c)}/M$,
  $\widehat{\widetilde{\tau}}_i^{(c)}=\hat{l}_i^{(c)}/N$.
\ENDFOR

\STATE \textbf{Second-stage rotation (fine ToA-AoA)}
\FOR{$i=1$ to $\hat{I}$}
  \STATE Build the compensated channel
  \[
  H_i^{(c)}(m,n)=H(m,n)\,
  \exp\!\Big(j2\pi\frac{\alpha}{N}mn\,\widehat{\widetilde{\phi}}_i^{(c)}\Big).
  \]
  \STATE Define small search sets $\mathcal{D}_\phi$ and $\mathcal{D}_\tau$.
  \FORALL{$(\Delta\widetilde{\phi},\Delta\widetilde{\tau})\in
           \mathcal{D}_\phi\times\mathcal{D}_\tau$}
     \STATE $\mathbf{R}_M=\mathrm{diag}(e^{j2\pi m\Delta\widetilde{\phi}})$,
            $\mathbf{R}_N=\mathrm{diag}(e^{j2\pi n\Delta\widetilde{\tau}})$.
     \STATE $\mathbf{f}_M=\mathbf{f}_M(\hat{k}_i^{(c)})$,
            $\mathbf{f}_N=\mathbf{f}_N(\hat{l}_i^{(c)})$.
     \STATE
     \[
     J_i(\Delta\widetilde{\phi},\Delta\widetilde{\tau})
     =
     \Big|
     \mathbf{f}_M^{H}\mathbf{R}_M
     \mathbf{H}_i^{(c)}
     \mathbf{R}_N\mathbf{f}_N
     \Big|.
     \]
  \ENDFOR
  \STATE $(\Delta\widehat{\widetilde{\phi}}_i,\Delta\widehat{\widetilde{\tau}}_i)
         =\arg\max J_i(\Delta\widetilde{\phi},\Delta\widetilde{\tau})$.
  \STATE
  $\widehat{\widetilde{\phi}}_i=
   \widehat{\widetilde{\phi}}_i^{(c)}+\Delta\widehat{\widetilde{\phi}}_i$,
  $\widehat{\widetilde{\tau}}_i=
   \widehat{\widetilde{\tau}}_i^{(c)}+\Delta\widehat{\widetilde{\tau}}_i$.
\ENDFOR

\STATE \textbf{Complex coefficient estimation}
\FOR{$i=1$ to $\hat{I}$}
  \STATE $\mathbf{R}_M=\mathrm{diag}(e^{j2\pi m(\widehat{\widetilde{\phi}}_i-\widehat{\widetilde{\phi}}_i^{(c)})})$,
         $\mathbf{R}_N=\mathrm{diag}(e^{j2\pi n(\widehat{\widetilde{\tau}}_i-\widehat{\widetilde{\tau}}_i^{(c)})})$.
  \STATE
  \[
  \widehat{\widetilde{\beta}}_i
  =
  \frac{1}{\sqrt{MN}}\,
  \mathbf{f}_M^{H}(\hat{k}_i^{(c)})
  \mathbf{R}_M\mathbf{H}_i^{(c)}\mathbf{R}_N
  \mathbf{f}_N(\hat{l}_i^{(c)}).
  \]
\ENDFOR
\end{algorithmic}
\normalsize
\end{algorithm}

\vspace{2 mm}
\noindent\textbf{\textcolor{blue}{Fine off-grid estimation via rotation:}}

\textcolor{blue}{%
Under the approximation in~\eqref{eq:Hc_narrowband_equiv}, the 2D IDFT of $H_i^{(c)}(m,n)$ within the $i$th cluster behaves as in the narrowband case of Section~\ref{sec_pre}, with a peak centered at the corrected coarse bin $(\hat{k}_i^{(c)},\hat{l}_i^{(c)})$. Let $\widehat{\widetilde{\phi}}_i^{(c)}=\hat{k}_i^{(c)}/M$ and $\widehat{\widetilde{\tau}}_i^{(c)}=\hat{l}_i^{(c)}/N$ denote the corresponding coarse normalized AoA and ToA signatures. We now apply a 2D rotation around these coarse values to estimate the fractional offsets.}

\textcolor{blue}{Define the spatial and frequency rotation matrices.}

\par\noindent\small
\color{blue}
\begin{align}
\mathbf{R}_M(\Delta\widetilde{\phi})
&\color{blue}\doteq
\mathrm{diag}\!\left(\e{j2\pi m\Delta\widetilde{\phi}}\right)_{m=0}^{M-1},
\label{eq:RM_def} \\
\mathbf{R}_N(\Delta\widetilde{\tau})
&\color{blue}\doteq
\mathrm{diag}\!\left(\e{j2\pi n\Delta\widetilde{\tau}}\right)_{n=0}^{N-1},
\label{eq:RN_def}
\end{align}
\normalsize

\textcolor{blue}{%
and the IDFT steering vectors}
\par\noindent\small
\begin{align}
\color{blue}
\mathbf{f}_M(\hat{k}_i^{(c)})
&\color{blue}\doteq
\frac{1}{\sqrt{M}}
\left[1, \ldots,
\e{j2\pi(M-1)\hat{k}_i^{(c)}/M}\right]^{T},
\label{eq:fM_def}\\
\color{blue}
\mathbf{f}_N(\hat{l}_i^{(c)})
&\color{blue}\doteq
\frac{1}{\sqrt{N}}
\left[1, \ldots,
\e{j2\pi(N-1)\hat{l}_i^{(c)}/N}\right]^{T}.
\label{eq:fN_def}
\end{align}
\normalsize

\textcolor{blue}{%
Let $\mathbf{H}_i^{(c)}$ denote the matrix formed from $H_i^{(c)}(m,n)$ restricted to the $i$th cluster. The second-stage fine-tuning problem is then formulated as the maximization of the rotation objective}
\par\noindent\small
\begin{align}
J_i(\Delta\widetilde{\phi},\Delta\widetilde{\tau})
=
\Big|
\mathbf{f}_M^{H}(\hat{k}_i^{(c)})\,
\mathbf{R}_M(\Delta\widetilde{\phi})\,
\mathbf{H}_i^{(c)}\,
\mathbf{R}_N(\Delta\widetilde{\tau})\,
\mathbf{f}_N(\hat{l}_i^{(c)})
\Big|.
\label{eq:J_second_stage}
\end{align}
\normalsize

\textcolor{blue}{%
In the noise-free case and under the narrowband approximation in \eqref{eq:Hc_narrowband_equiv}, the objective $J_i(\Delta\widetilde{\phi},\Delta\widetilde{\tau})$ achieves its maximum at the true residual offsets
$\Delta\widetilde{\phi}_i^{\star}
=
\widetilde{\phi}_i-\widehat{\widetilde{\phi}}_i^{(c)}$ and
$\Delta\widetilde{\tau}_i^{\star}
=
\widetilde{\tau}_i-\widehat{\widetilde{\tau}}_i^{(c)}$, so that the final fine off-grid
estimates are}
\par\noindent\small
\begin{align}
\widehat{\widetilde{\phi}}_i
=
\widehat{\widetilde{\phi}}_i^{(c)}+\Delta\widetilde{\phi}_i^{\star},\qquad
\widehat{\widetilde{\tau}}_i
=
\widehat{\widetilde{\tau}}_i^{(c)}+\Delta\widetilde{\tau}_i^{\star}.
\label{eq:final_fine_estimates}
\end{align}
\normalsize
\textcolor{blue}{%
This completes the second-stage rotation-based fine-tuning step.}

\newpage
\noindent\textbf{\textcolor{blue}{Complex path coefficient estimation:}}

\textcolor{blue}{Once the refined estimates 
$\widehat{\widetilde{\phi}}_i$ and $\widehat{\widetilde{\tau}}_i$ are obtained via
\eqref{eq:final_fine_estimates}, the complex gain $\widetilde{\beta}_i$
is estimated from the compensated channel $\mathbf{H}_i^{(c)}$ defined in
\eqref{eq:Hc_def}. We define the residual offsets as}
\par\noindent\small
\begin{align}
\textcolor{blue}{
\Delta\widehat{\widetilde{\phi}}_i =
\widehat{\widetilde{\phi}}_i -
\widehat{\widetilde{\phi}}_i^{(c)},\qquad
\Delta\widehat{\widetilde{\tau}}_i =
\widehat{\widetilde{\tau}}_i -
\widehat{\widetilde{\tau}}_i^{(c)}}.
\label{eq:fine_residuals_fixed}
\end{align}
\normalsize

\textcolor{blue}{The corresponding fine-rotation matrices are given by}
\par\noindent\small
\begin{align}
\mathbf{R}_M\!\left(\Delta\widehat{\widetilde{\phi}}_i\right)
&=
\mathrm{diag}\!\left(
\e{j2\pi m\,\Delta\widehat{\widetilde{\phi}}_i}
\right)_{m=0}^{M-1}, \label{eq:RM_fixed}\\
\mathbf{R}_N\!\left(\Delta\widehat{\widetilde{\tau}}_i\right)
&=
\mathrm{diag}\!\left(
\e{j2\pi n\,\Delta\widehat{\widetilde{\tau}}_i}
\right)_{n=0}^{N-1}. \label{eq:RN_fixed}
\end{align}
\normalsize

\textcolor{blue}{Let $\mathbf{f}_M(\hat{k}_i^{(c)})$ and 
$\mathbf{f}_N(\hat{l}_i^{(c)})$ denote the $M$- and $N$-point normalized IDFT
steering vectors. The complex gain is estimated via}
\par\noindent\small
\begin{align}
\textcolor{blue}{
\widehat{\widetilde{\beta}}_i
=
\frac{1}{\sqrt{MN}}\,
\mathbf{f}_M^{H}(\hat{k}_i^{(c)})
\,
\mathbf{R}_M\!\left(\Delta\widehat{\widetilde{\phi}}_i\right)
\,
\mathbf{H}_i^{(c)}
\,
\mathbf{R}_N\!\left(\Delta\widehat{\widetilde{\tau}}_i\right)
\,
\mathbf{f}_N(\hat{l}_i^{(c)})}.
\label{eq:beta_est_final_fixed}
\end{align}
\normalsize

\textcolor{blue}{Because the residual dispersion is bounded (\eqref{eq:residual_phase_bound}), the neighboring path contributions 
inside the refined local region are negligible, making \eqref{eq:beta_est_final_fixed} a consistent projection-based estimator in the noise-free regime.} We write the concrete mathematical stepwise procedure as given in Algorithm \ref{alg:2stage_swe}.

We can also evaluate the actual AoA-ToA from the estimated normalized AoA-ToA as
\par\noindent\small
\begin{align}
    \hat{\phi}_i = arcsin\left(\frac{\lambda}{d}\hat{\hat{\Tilde{\phi}}}_i\right),
    \hat{\tau}_i = \hat{\hat{\Tilde{\tau}}}_i\frac{N}{\alpha f_c}.
\end{align}
\normalsize

\begin{table*}[h]
\centering
\caption{Complexity comparison of joint DoA-ToA estimation methods}
\label{tab:complexity_compare}
\small
\resizebox{\textwidth}{!}{
\begin{tabular}{|c|c|c|}
\hline
\textbf{Method} &
\textbf{Complexity Order} &
\textbf{Notes} \\ \hline

DFT-only (coarse) &
$\mathcal{O}(MN\log MN)$ &
Grid-based resolution (Narrowband systems) \\ \hline

Direct rotation (Wang \textit{et al.} \cite{wang2018spatial}) &
$\mathcal{O}(IR_MR_NMN)$ &
SWE not compensated $\Rightarrow$ degraded coarse estimation \\ \hline

OMP-based Sparse Recovery &
$\mathcal{O}(IKMN+K^3)$ &
High dictionary correlation and iterative support update \\ \hline

MUSIC/ESPRIT &
$\mathcal{O}(M^3+N^3)$ &
Covariance estimation and full eigen-decomposition \\ \hline

Proposed Two-Stage Rotation (Wideband) &
$\mathcal{O}\!\big(MN\log MN + I|\mathcal{M}||\mathcal{N}|\log(MN) + IR_MR_NMN\big)$ &
Handles coarse-bin shift and leakage due to SWE with near-linear grid scaling \\ \hline

Proposed (Narrowband - special case) &
$\mathcal{O}\!\big(MN\log MN + IR_MR_NMN\big)$ &
$|\mathcal{M}|=|\mathcal{N}|=1$ $\Rightarrow$ direct reduction to spatial narrowband case \\ \hline

\end{tabular}
}
\normalsize
\vspace{-3mm}
\end{table*}

\subsection{Computational Complexity}

\textcolor{blue}{
The initial 2D-DFT of the $M\times N$ space–frequency response requires $\mathcal{O}(MN\log MN)$ operations, and the subsequent peak detection over $MN$ grid points incurs $\mathcal{O}(MN)$ cost, dominated by the former. In the first-stage rotation, spatial wideband compensation and local coarse-bin correction are performed within a bounded neighborhood $\mathcal{M}\times\mathcal{N}$ around each detected cluster, leading to a total complexity of $\mathcal{O}\big(I|\mathcal{M}||\mathcal{N}|\log(MN)\big)$ for $I$ identified paths. In the second-stage rotation, a refined $R_M\times R_N$ search per path entails correlation operations over the entire $MN$ grid, resulting in $\mathcal{O}\big(IR_MR_NMN\big)$ complexity. The subsequent estimation of the $I$ complex path gains involves $\mathcal{O}(IMN)$ operations and is dominated by the fine-rotation step. Hence, the overall complexity of the proposed method is $\mathcal{O}\!\big(MN\log MN + I|\mathcal{M}||\mathcal{N}|\log(MN) + IR_MR_NMN\big)$, which reduces to $\mathcal{O}\!\big(MN\log MN + IR_MR_NMN\big)$ for spatially narrowband systems where $|\mathcal{M}| = |\mathcal{N}| = 1$. The computational complexity comparison of different methods is summarized in Table~\ref{tab:complexity_compare}}.

\section{Results}
\label{sec_results}

\begin{figure*}[htbp]
  \centering
  \begin{subfigure}[b]{0.23\linewidth}
    \centering
    \includegraphics[width=\linewidth]{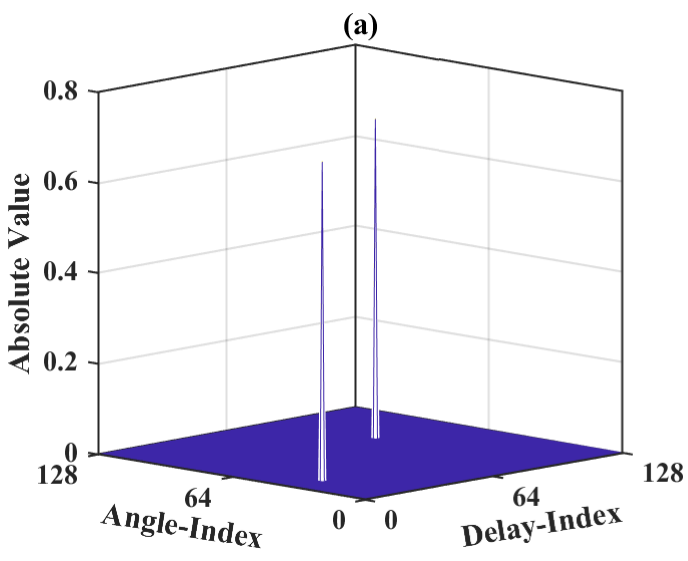}
  \end{subfigure}
  \hfill
  \begin{subfigure}[b]{0.23\linewidth}
    \centering
    \includegraphics[width=\linewidth]{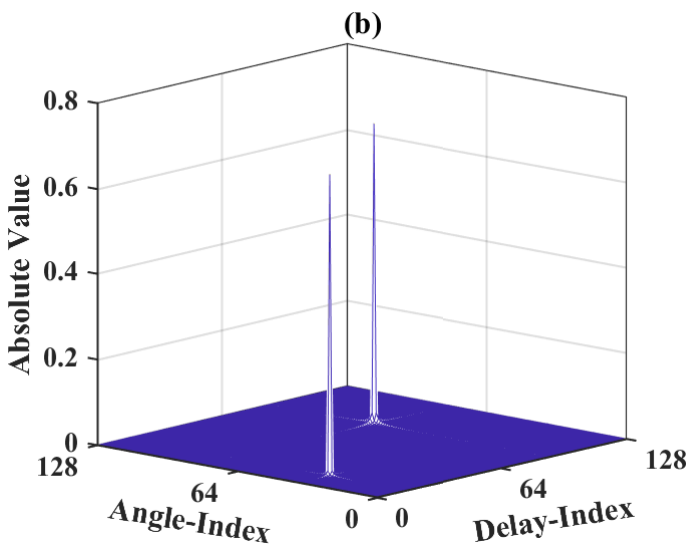}
  \end{subfigure}
  \hfill
  \begin{subfigure}[b]{0.23\linewidth}
    \centering
    \includegraphics[width=\linewidth]{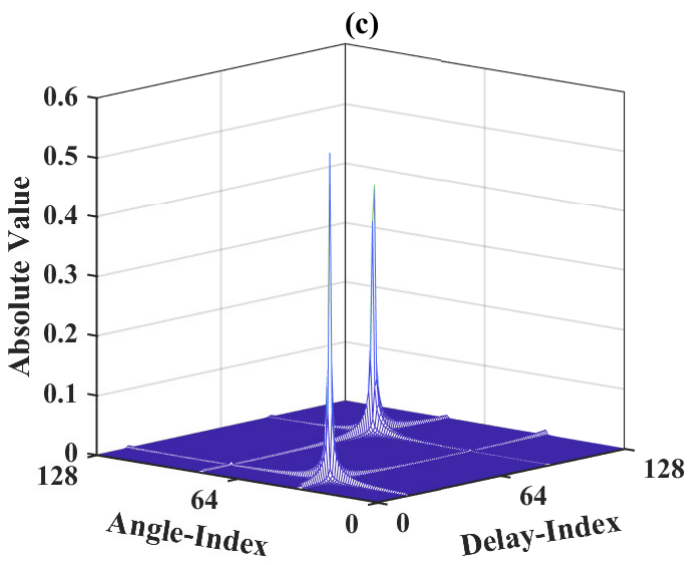}
  \end{subfigure}
  \hfill
  \begin{subfigure}[b]{0.23\linewidth}
    \centering
    \includegraphics[width=\linewidth]{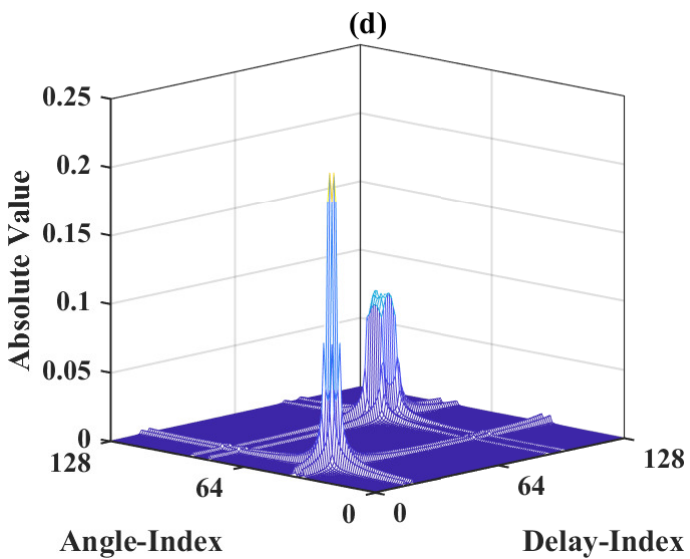}
  \end{subfigure}
  \caption{Delay-Angle CIR for (a) spatial narrowband model with integer signature  (b) spatial wideband model ($\alpha = 0.01$) with integer signature (c) spatial wideband model ($\alpha = 0.01$) with non-integer signature (d) spatial wideband model ($\alpha = 0.1$) with non-integer signature.}
  \label{fig:spatial_wideband_effect}
\end{figure*}

\textcolor{blue}{
In this section, we evaluate the performance of the proposed signature estimation algorithm through Monte–Carlo simulations. Unless stated otherwise, we 
consider a MIMO–OFDM setup with $M=N=128$, $\alpha = 0.1$, $f_c = 73$ GHz and two propagation paths with equal complex gains 
$[0.5+0.5j,\,0.5+0.5j]$. To clearly illustrate the respective impacts of SWE and spectral leakage under different delay–angle 
configurations, Fig.~\ref{fig:spatial_wideband_effect} shows the normalized 2D delay–angle channel magnitudes for four representative cases.}
\textcolor{blue}{%
Fig.~\ref{fig:spatial_wideband_effect}(a) shows the spatial \textit{narrowband} case ($\alpha=0$) with integer grid-aligned signatures 
$(35/M,15/N)$ and $(80/M,88/N)$. Both paths appear as isolated impulses with zero leakage, validating Lemma~\ref{lemma_1}.  
When a very small wideband factor is introduced ($\alpha=0.01$) while retaining integer signatures, mild spreading along the delay axis appears 
in Fig.~\ref{fig:spatial_wideband_effect}(b), yet the maximum power still coincides with the correct coarse bins.} \textcolor{blue}{%
In Fig.~\ref{fig:spatial_wideband_effect}(c), we keep $\alpha=0.01$ but introduce off-grid signatures $(35.25/M,15.25/N)$ and $(80.25/M,88.50/N)$. 
Here, the Dirichlet kernels generate symmetric spectral leakage around the nearest DFT bins, as predicted by Section~\ref{sec_pre}.  
Finally, in Fig.~\ref{fig:spatial_wideband_effect}(d), a larger wideband factor ($\alpha=0.1$) causes coupled angular–delay spreading governed by Theorem~\ref{thm:SWE_spreading}, leading to significant coarse-bin displacement in addition to leakage. This highlights the necessity of the proposed two-stage rotation scheme to correctly recover both coarse and fractional signatures under SWE.}

\begin{figure}[h]
  \centering
  \begin{subfigure}[b]{0.45\linewidth}
    \centering
    \includegraphics[width=\linewidth]{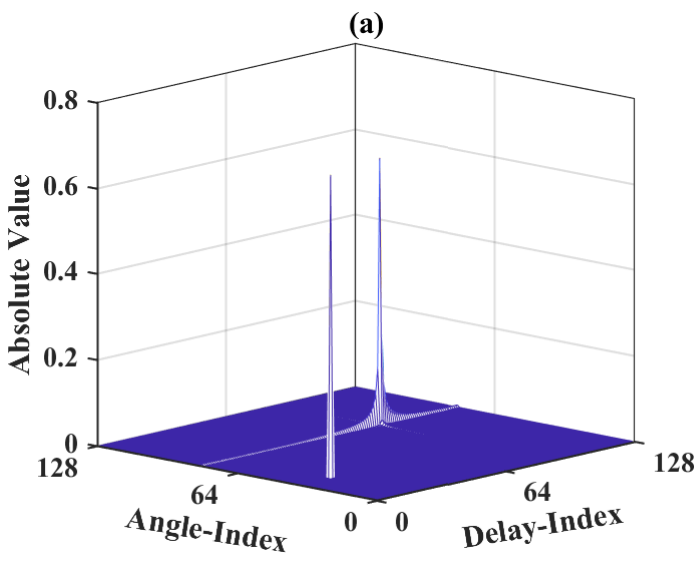}
  \end{subfigure}
  \hfill
  \begin{subfigure}[b]{0.45\linewidth}
    \centering
    \includegraphics[width=\linewidth]{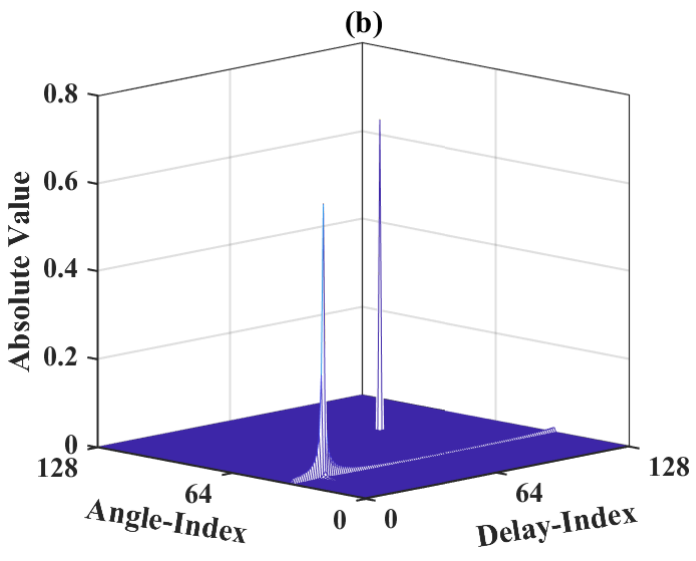}
  \end{subfigure}
  \caption{Rotation-based ToA-AoA estimation for spatial narrowband system $(\alpha = 0.01)$ (a) path-1 at $(35.25/M, 15.25/N)$ (b) path-2 at $(80.25/M,88.50/N)$}.
  \label{fig:one-stage rotation estimation}
\end{figure}

\vspace{0.5cm}
\textbf{1) Numerical example:} In solution, we first show the effectiveness of the rotation-based technique in the spatial narrowband system for fine-tuning around the estimated coarse signature. In Fig. \ref{fig:one-stage rotation estimation}, we see that with $R_M, R_N=5$, we can exactly localize the paths, and the estimated output for path-1 is $(35.25/M, 15.25/N)$ and for path-2 is $(80.25/M,88.50/N)$. The correspondingly estimated path coefficients are $(0.50+0.49i)$ and $(0.50+0.49i)$. The small difference in estimated path coefficients is due to the use of the spatial wideband model ($\alpha \neq 0)$.

\begin{figure}[h]
  \centering
  \begin{subfigure}[b]{0.48\linewidth}
    \centering
    \includegraphics[width=\linewidth]{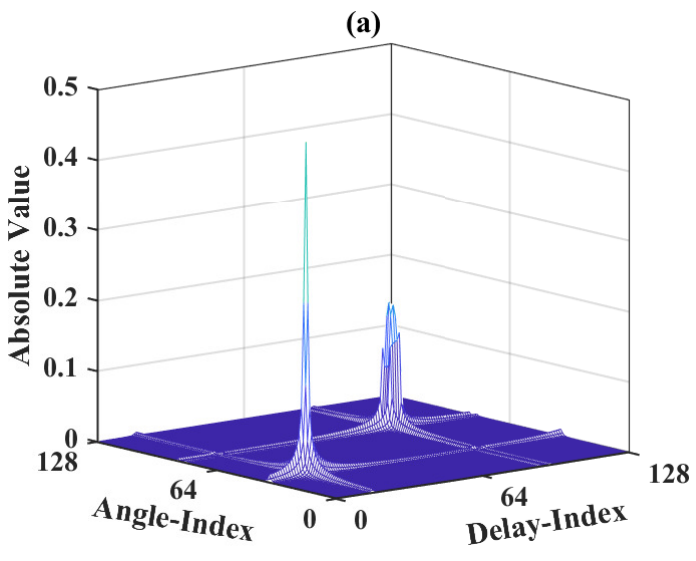}
  \end{subfigure}
  \hfill
  \begin{subfigure}[b]{0.48\linewidth}
    \centering
    \includegraphics[width=\linewidth]{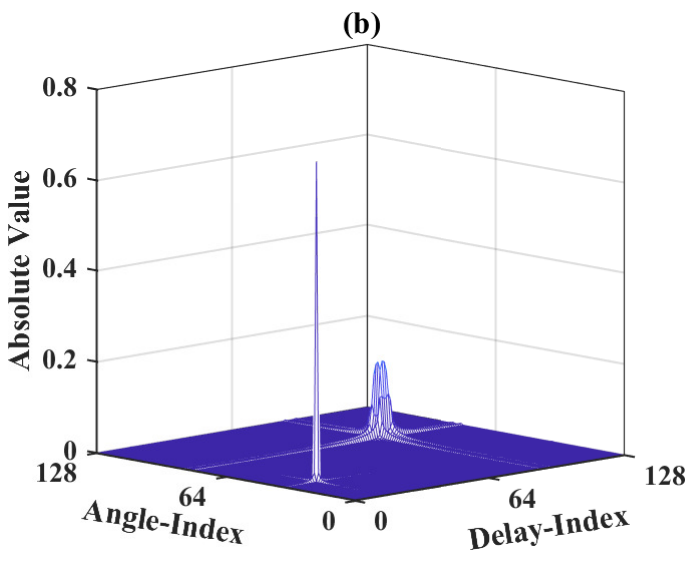}
  \end{subfigure}
  \caption{Rotation-based ToA-AoA estimation for spatial wideband system $(\alpha = 0.1)$ (a) path-1 with one-stage rotation (b) path-1 with two-stage rotation}
  \label{fig:two-stage rotation path-1}
\end{figure}

\begin{figure}[h]
  \centering
  \begin{subfigure}[b]{0.48\linewidth}
    \centering
    \includegraphics[width=\linewidth]{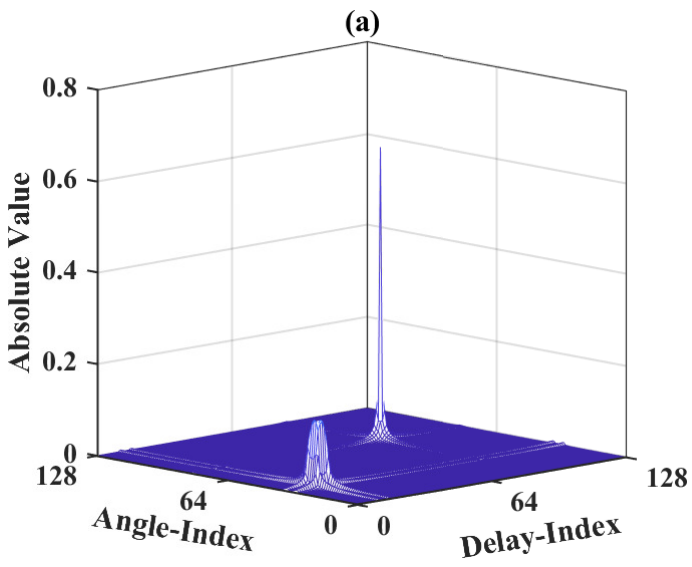}
  \end{subfigure}
  \hfill
  \begin{subfigure}[b]{0.48\linewidth}
    \centering
    \includegraphics[width=\linewidth]{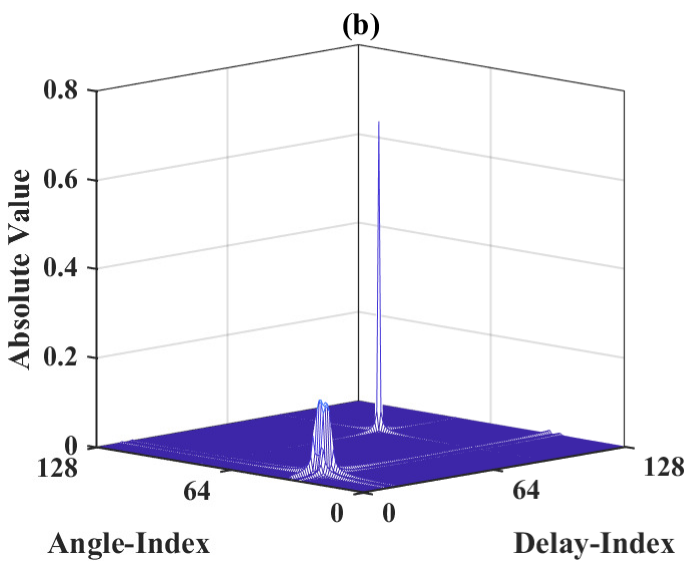}
  \end{subfigure}
  \caption{Rotation-based ToA-AoA estimation for spatial wideband system $(\alpha = 0.1)$ (a) path-2 with one-stage rotation (b) path-2 with two-stage rotation.}
  \label{fig:two-stage rotation path-2}
\end{figure}

Further, we discuss the spatial wideband effect for each path. If we directly pick the peaks of every cluster as the coarse bin, i.e. $(36,17)$ for path-1 and $(87,92)$ for path-2, we can see from Figs. \ref{fig:two-stage rotation path-1}(a) and \ref{fig:two-stage rotation path-2}(a) that due to these incorrect coarse, the wideband effect cannot be conjugated, and the wideband spread will persist. However, with the neighborhood $|\mathcal{M}|,|\mathcal{N}| = 13$, in the first stage of rotation of the proposed method, we get the correct coarse bins $(35,15)$ and $(80,88)$. The wideband effect can be compensated here up to the integer values, as shown in Figs. \ref{fig:two-stage rotation path-1}(b) and \ref{fig:two-stage rotation path-2}(b). The estimated fine-tuned signature in the second stage of the proposed method, around the corrected coarse bin with $R_M,R_N = 5$ are $(35.25/M,15.25/N)$ and $(80.25/M,88.50/N)$. Also, the estimated complex path coefficient is $0.57+0.40i$ with a small error persisting due to the fractional parts of the input signature.


\begin{figure}
    \centering
    \includegraphics[width=0.8\linewidth]{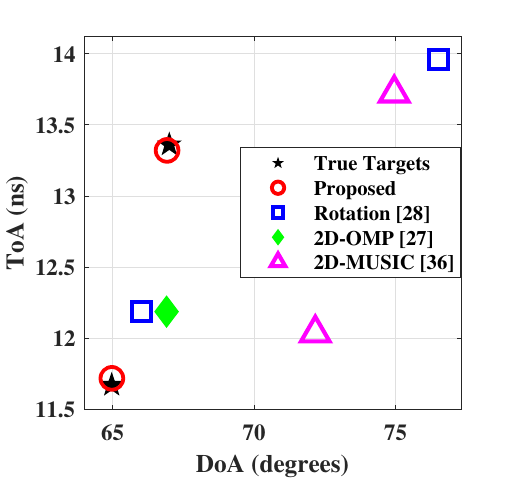}
    \caption{Close-proximity scatter localization in the DoA-ToA domain.}
    \label{fig:scatter_proximity}
\end{figure}

\textbf{2) Scatter proximity:}
Fig.~\ref{fig:scatter_proximity} demonstrates the ability of the proposed framework to resolve closely spaced scatters in the joint delay-angle domain. Two targets are considered with true parameter pairs $(\theta,\tau) = (65^\circ,\,11.67~\text{ns})$ and $(67^\circ,\,13.33~\text{ns})$. The proposed two-stage rotation-based method accurately localizes both scatters at $(64.99^\circ,\,11.67~\text{ns})$ and $(66.97^\circ,\,13.33~\text{ns})$, achieving near-perfect estimation despite the small separation in both angular and temporal dimensions. In contrast, the 2D-OMP method detects only a single scatter, indicating its inability to resolve closely spaced targets under severe dictionary coherence and grid mismatch effects. The direct (single-stage) rotation technique identifies two targets; however, the estimated locations suffer from noticeable errors due to coarse bin shifting, which is not adequately compensated in the absence of a refinement stage. Furthermore, the 2D-MUSIC method fails to localize the true scatter positions, primarily due to significant model mismatch arising from the use of a conventional spatial narrowband assumption in the presence of spatial wideband effects. These results clearly highlight the superior resolution capability and robustness of the proposed method for closely spaced scatter localization in spatially wideband scenarios.

\begin{figure*}
    
  \centering
  \begin{subfigure}[b]{0.32\linewidth}
    \centering
    \includegraphics[width=\linewidth]{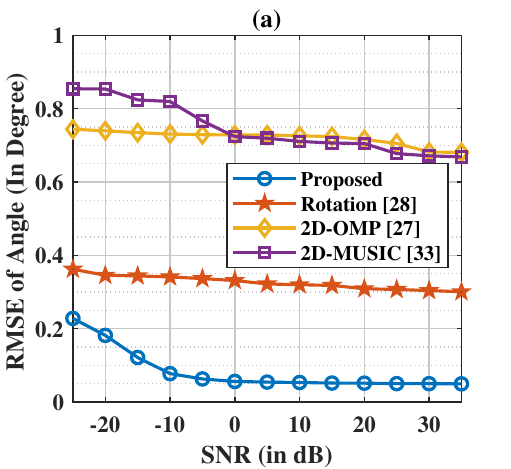}
  \end{subfigure}
  \hfill
    \centering
  \begin{subfigure}[b]{0.32\linewidth}
    \centering
    \includegraphics[width=\linewidth]{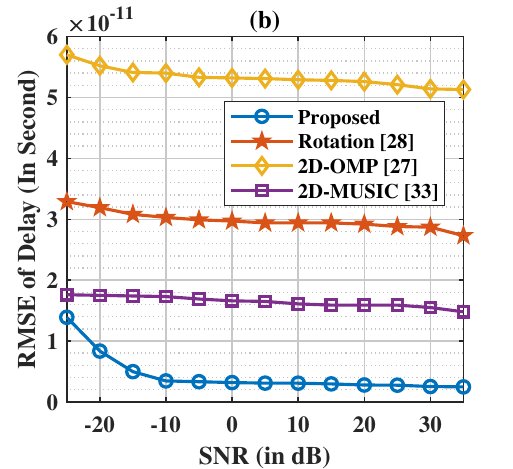}
  \end{subfigure}
  \hfill
  \begin{subfigure}[b]{0.32\linewidth}
    \centering
    \includegraphics[width=\linewidth]{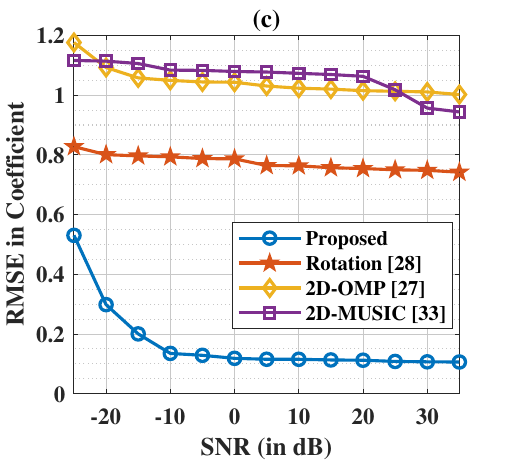}
  \end{subfigure}
  \caption{RMSE in (a) AoA, (b) ToA, and (c) complex coefficient for different methods at high BW ($\alpha = 0.1$) and $M,N = 128$.}
  \label{fig:rmse}
\end{figure*}

For numerical simulation, we considered $K=5$ targets with uniformly distributed ToAs and AoAs. The target gains were generated as $\Tilde{\beta_{k}}=\exp{(j\psi_{k})}$ with $\psi_{k}\sim\mathcal{U}[0,2\pi)$ while the noise $w_{m,n}\sim \mathcal{CN}(\mathbf{0},\sigma^{2}I)$, i.i.d. across all space-frequency samples. We evaluate the detection and estimation performance of our proposed method in comparison to the direct rotation approach \cite{wang2018spatial}, as well as the 2D-OMP \cite{rai2025low} and 2D-MUSIC \cite{belfiori20122d} techniques.

A detected target is classified as a "hit" if its estimated ToA and AoA fall within the resolution of the true target parameters; otherwise, it is considered a false alarm. Thresholds were adjusted to ensure a constant false-alarm rate across all SNRs, while the root mean squared error (RMSE) in estimation was calculated for targets identified as hits. Unless stated otherwise, the rates and RMSE values were averaged over 300 independent trials.
We define the RMSE for the correctly estimated signatures as
\par\noindent\small
\begin{align*}
    RMSE_{\phi} \triangleq \sqrt{\frac{1}{\hat{I}} \sum\limits_{i=1}^{\hat{I}}(\hat{\phi_i}-\phi_i)^{2}}, 
    RMSE_{\tau} \triangleq \sqrt{\frac{1}{\hat{I}} \sum\limits_{i=1}^{\hat{I}}(\hat{\tau_i}-\tau_i)^{2}} .
\end{align*}

\begin{align*}
    RMSE_{\Tilde{\beta}} \triangleq \sqrt{\frac{1}{\hat{I}} \sum\limits_{i=1}^{\hat{I}} \norm{(\hat{\Tilde{\beta}}_i-\Tilde{\beta}_i)}^2}.
    \label{eq:mse_beta}
\end{align*}
\normalsize

\begin{figure}
  \centering
  \begin{subfigure}[b]{0.48\columnwidth}
    \centering
    \includegraphics[width=\linewidth]{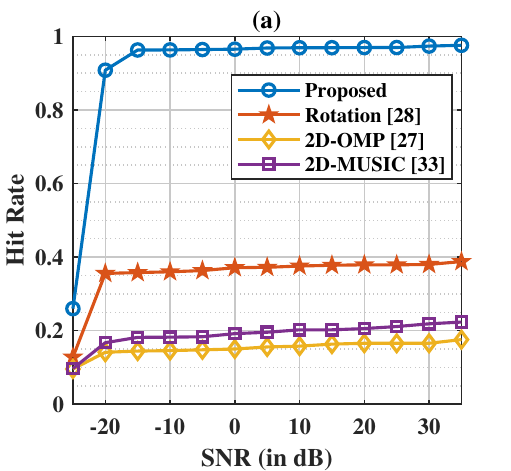}
  \end{subfigure}
  \hfill
  \begin{subfigure}[b]{0.48\columnwidth}
    \centering
    \includegraphics[width=\linewidth]{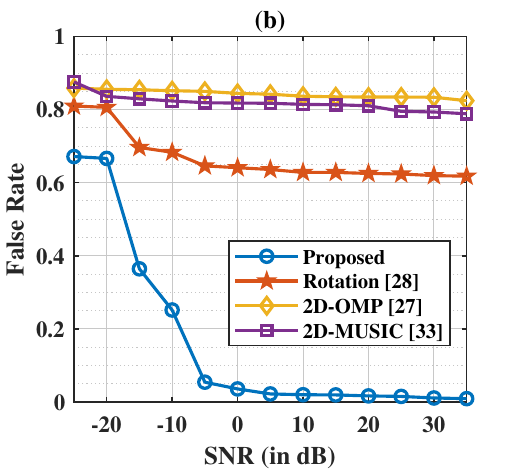}
  \end{subfigure}
  \caption{(a) Hit rate and (b) false rate of different methods at high BW ($\alpha = 0.1$) and $M,N = 128$ w.r.t. SNR variation.}
  
  \label{fig:hit_rate}
\end{figure}

\textbf{3) Performance Comparison with other methods:} We show the scatter detection performance in terms of hit rate and false rate of different methods for SNRs ranging from $-25$dB to $35$dB in Fig. \ref{fig:hit_rate}. The corresponding RMSE in ToA, AoA and complex coefficient is shown in Fig. \ref{fig:rmse}. It can be observed from Fig. \ref{fig:hit_rate}(a) that the 2D-OMP and 2D-MUSIC methods perform with a small hit rate ($\approx 0.2$) and large false rate ($\approx 0.8$ ) since these methods strictly work for narrowband signal models. Moreover, the direct rotation technique also yields a limited hit rate up to $0.4$ and still high false rate of around $0.6$. However, our proposed two-stage rotation method yields the hit rate as high as $0.98$. Similarly, the false rate is reduced to as low as $0.01$ for high SNR values as shown in Fig. \ref{fig:hit_rate}(b). Hence, from Fig. \ref{fig:hit_rate}, we see that the two-stage rotation is essential for high BW ($\alpha = 0.1$) to correctly detect the scatters.

Furthermore, the estimation error in all three parameters is the lowest for the proposed two-stage rotation method, as shown in Fig. \ref{fig:rmse}. The direct rotation method still gives inferior RMSEs as compared to the proposed method due to the inability to compensate for the spatial wideband effect fully. Moreover, the 2D-MUSIC and 2D-OMP methods, being model specific yields high RMSEs as depicted in Fig. \ref{fig:rmse}. 

\begin{figure}
  \centering
  \begin{subfigure}[b]{0.48\columnwidth}
    \centering
    \includegraphics[width=\linewidth]{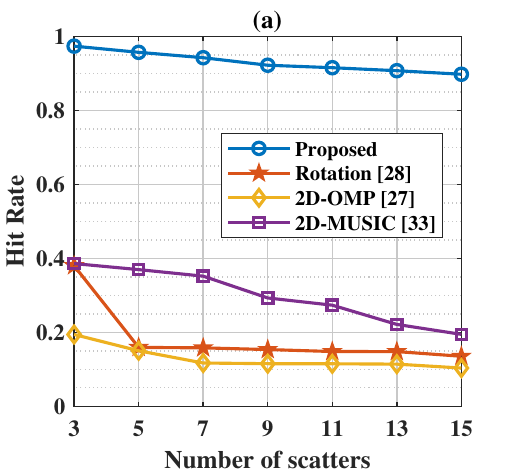}
  \end{subfigure}
  \hfill
  \begin{subfigure}[b]{0.48\columnwidth}
    \centering
    \includegraphics[width=\linewidth]{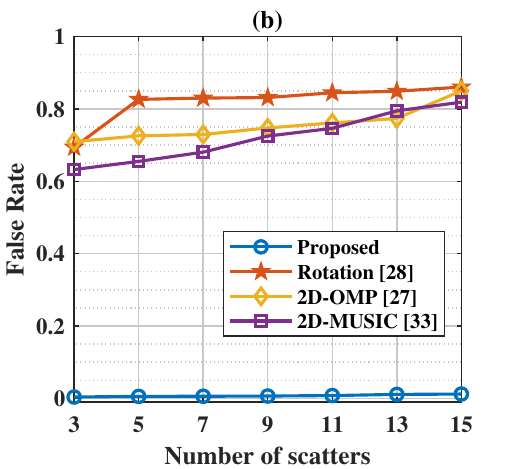}
  \end{subfigure}
  \caption{(a) Hit rate and (b) False rate at high BW ($\alpha = 0.1$) and $M,N = 128$ w.r.t. no. of scatters variation.}
  \label{fig:hit_rate_01}
\end{figure}

\textbf{4) Performance comparison in the presence of variable number of scatters:} We show the scalability of our proposed method with respect to the increasing number of targets in Fig. \ref{fig:hit_rate_01}. The hit rate of the proposed two-stage rotation method is high and decreases quite slowly with the increase in the number of scatters. In the presence of a high number of scatters, there is more chance that the two or more paths overlap and hence the less number of clusters are detected. This causes a decrease in the hit rate. However, as we can verify from Fig. \ref{fig:hit_rate_01}(b), the false rate is still kept low with the proposed method. Also, the other computational techniques yield consistently poor hit rate and false rate as depicted in Fig. \ref{fig:hit_rate_01}.

\begin{figure}
  \centering
  \begin{subfigure}[b]{0.48\columnwidth}
    \centering
    \includegraphics[width=\linewidth]{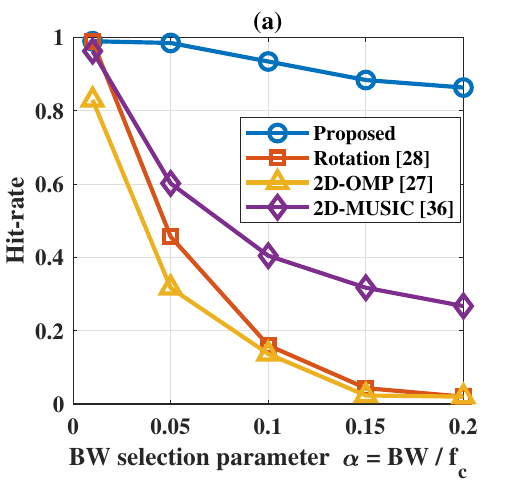}
  \end{subfigure}
  \hfill
  \begin{subfigure}[b]{0.48\columnwidth}
    \centering
    \includegraphics[width=\linewidth]{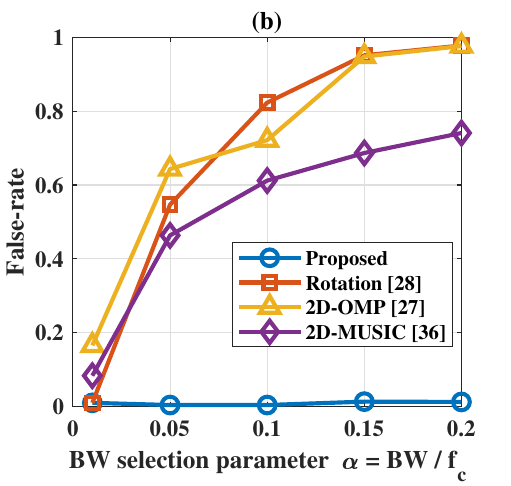}
  \end{subfigure}
  \caption{(a) Hit rate and (b) False rate w.r.t. BW variation.}
  \label{fig:hit_rate_a_varied}
\end{figure}

\vspace{0.5cm}
\textbf{5) Performance comparison with variable bandwidth:} Fig.~\ref{fig:hit_rate_a_varied} illustrates the variation of hit rate and false rate as a function of the normalized bandwidth parameter $\alpha = \mathrm{BW}/f_c$, which directly governs the severity of spatial wideband effects. As shown in Fig.~\ref{fig:hit_rate_a_varied}(a), the proposed two-stage rotation-based method maintains a consistently high hit rate across the entire range of $\alpha$, exhibiting only a gradual degradation as the bandwidth increases. This behavior indicates the robustness of the proposed framework against spatial wideband distortions, owing to its explicit modeling of wideband effects and the subsequent fine parameter refinement stage. In contrast, the direct (single-stage) rotation method experiences a rapid decline in hit rate with increasing $\alpha$, as coarse bin shifting and residual wideband errors are not adequately compensated. The 2D-OMP method suffers an even more pronounced performance degradation due to increased dictionary coherence and severe grid mismatch at higher bandwidths, leading to missed detections. Similarly, the 2D-MUSIC method exhibits a monotonic reduction in hit rate as $\alpha$ increases, primarily due to model mismatch arising from the use of a conventional spatial narrowband assumption in a spatially wideband regime.
    
The corresponding false-rate trends are shown in Fig.~\ref{fig:hit_rate_a_varied}(b). The proposed method consistently achieves a near-zero false rate across all values of $a$, demonstrating its strong discrimination capability and robustness to wideband-induced artifacts. In contrast, the false rates of the baseline methods increase sharply with $\alpha$. The direct rotation and 2D-OMP methods approach near-unit false rates for moderate-to-large values of $\alpha$, reflecting the emergence of spurious detections caused by dictionary leakage and unmodeled wideband effects. Although the 2D-MUSIC method exhibits a relatively slower increase in false rate compared to sparse recovery approaches, it nevertheless suffers from substantial false alarms at higher bandwidths due to persistent model mismatch. 

\vspace{0.5cm}
\textbf{6) Performance comparison with variable grid size:} 

\begin{figure}
  \centering
  \begin{subfigure}[b]{0.48\columnwidth}
    \centering
    \includegraphics[width=\linewidth]{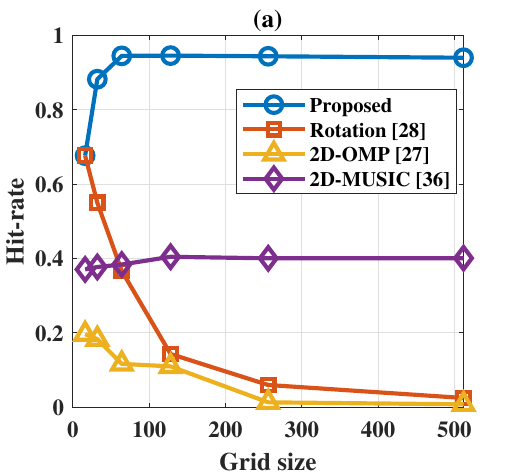}
  \end{subfigure}
  \hfill
  \begin{subfigure}[b]{0.48\columnwidth}
    \centering
    \includegraphics[width=\linewidth]{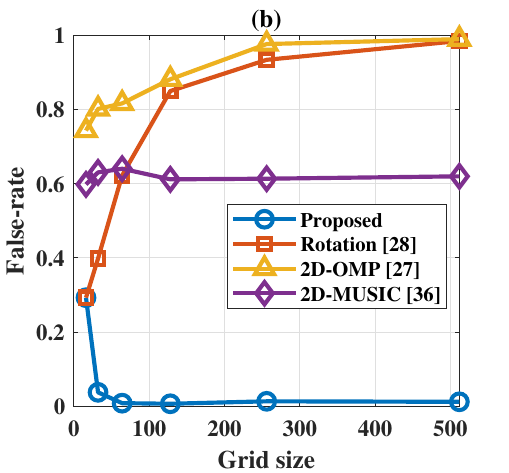}
  \end{subfigure}
  \caption{(a) Hit rate and (b) False rate w.r.t. grid-size variation.}
  \label{fig:hit_rate_nx_ny_varied}
\end{figure}

Fig.~\ref{fig:hit_rate_nx_ny_varied} depicts the variation of hit rate and false rate with respect to the measurement grid size, defined by the antenna-subcarrier product (i.e., $16\times16$ to $512\times512$). As shown in Fig.~\ref{fig:hit_rate_nx_ny_varied}(a), the proposed two-stage rotation-based method achieves a consistently high hit rate that rapidly saturates with increasing grid resolution. Even for moderate grid sizes, the proposed method attains near-optimal detection performance, indicating that the subsequent refinement stage effectively mitigates grid discretization errors and renders the performance largely insensitive to further grid refinement. In contrast, the direct (single-stage) rotation method exhibits a monotonic degradation in hit rate as the grid size increases. This behavior can be attributed to coarse bin shifting and the absence of a coarse correction mechanism, which leads to increasing estimation bias and missed detections. The 2D-OMP method suffers from a severe performance collapse with grid refinement, detecting very few targets for large grid sizes due to increased dictionary coherence and strong grid mismatch effects that hinder reliable sparse recovery. The 2D-MUSIC method demonstrates a relatively stable but limited hit rate across all grid sizes, reflecting its restricted resolution capability under spatial wideband conditions and its reliance on a narrowband signal model.
    
The corresponding false-rate trends are illustrated in Fig.~\ref{fig:hit_rate_nx_ny_varied}(b). The proposed method consistently maintains a near-zero false rate across all grid sizes, highlighting its strong robustness against spurious detections and its ability to reliably discriminate true targets even under dense grid configurations. In contrast, the false rates of the direct rotation and 2D-OMP methods increase rapidly with grid refinement, approaching unity for large grid sizes. This is primarily due to increased dictionary coherence, spectral leakage, and unmodeled wideband effects, which result in a proliferation of false alarms. The 2D-MUSIC method exhibits a comparatively moderate and nearly grid-invariant false rate; however, this apparent stability is achieved at the cost of limited detection accuracy, as reflected by its consistently lower hit rate. 

\section{Conclusion} 
\textcolor{blue}{
In this work, we have proposed a two-stage rotation-based algorithm for enhanced signature estimation of an immobile radio scenario.
We have identified that due to the spatial wideband effect into consideration, the path power spreads in the delay-angle domain, and correspondingly, the peaks for each path get shifted from the actual coarse bin. Further, we prove that the error incurred due to the bin shifting cannot be handled unless the two-stage algorithm is used. We establish via numerical simulation that the error in scatter detection for spatial wideband systems $(\alpha \geq 0.05)$ with $(M, N \geq 128)$ is considerably high by direct rotation and can never be recovered, irrespective of the SNR. However, with our proposed algorithm, the error reduces drastically in ToA, AoA, and complex coefficient estimation. Similarly, the hit rate of the proposed method increases with SNR while keeping a low false alarm rate. We further demonstrate that spatial narrowband ($\alpha = 0.01$) can be treated as a specific instance of our proposed spatial wideband solution.}

\textcolor{blue}{
The proposed framework assumes that the delay–angle clusters corresponding to different propagation paths remain well separated. In practical spatial wideband systems, closely spaced scatterers in the AoA–ToA domain may lead to partial or complete cluster overlap due to SWE-induced spreading and peak migration, making reliable path separation challenging. Relaxing this non-overlapping assumption and developing robust methods for resolving overlapping signatures constitutes an important direction for future work. In addition, for XL-MIMO configurations, near-field effects become prominent at short ranges, and their joint modeling and mitigation together with SWE represent another promising avenue for future investigation.}

\appendices
\section{Spatial wideband MIMO-FMCW radar system}
\label{app_radar_model}
We consider a MIMO FMCW radar where each transmit element emits a Linear Frequency Modulated (LFM) waveform. The baseband chirp transmitted from a generic antenna can be represented as
\par\noindent\small
\begin{align*}
x(t) = \exp\left(j2\pi\left(f_{0}t+\tfrac{\kappa}{2}t^{2}\right)\right), \quad 0\leq t\leq T_c,
\end{align*}
\normalsize
where $f_{0}$ denotes the carrier frequency, $\kappa$ is the chirp rate (slope), and $T_c$ is the chirp duration.

Assume a scene with $L$ point targets, each modeled as a non-fluctuating scatter. The $\ell$-th target is characterized by its range $r_{\ell}$, radial velocity $v_{\ell}$, and azimuth angle $\phi_{\ell}$. Orthogonality across different transmitters is achieved using time-division multiplexing (TDM), such that the signal at the $p$-th receiver due to the $m$-th transmitter can be written as
\par\noindent\small
\begin{align*}
z_{m,p}(t)=\sum_{\ell=1}^{L} b_{\ell}\, x(t-\tau^{\ell}_{m,p}), \quad 0\leq t\leq T_c,
\end{align*}
\normalsize
where $b_{\ell}$ is the reflection coefficient proportional to the target radar cross-section (RCS), and $\tau^{\ell}_{m,p}$ is the total propagation delay.

For notational simplicity, the virtual array formed by the MIMO configuration is indexed by $u=0,1,\ldots,U-1$, with uniform inter-element spacing $d$. The received signal at the $u$-th virtual element is expressed as
\par\noindent\small
\begin{align}
z_{u}(t)=\sum_{\ell=1}^{L} b_{\ell}\, x(t-\tau^{\ell}_{u}), \quad 0\leq t\leq T_c,
\end{align}
\normalsize
with the composite delay decomposed into
\par\noindent\small
\begin{align}
\tau^{\ell}_{u}=\tau^{\ell}_{r}+\tau^{\ell}_{d}+\tau^{\ell}_{u,\phi}, 
\label{eq:delay_parts}
\end{align}
\normalsize
where $\tau^{\ell}_{r}=2r_{\ell}/c$ is the round-trip propagation delay, $\tau^{\ell}_{d}=2v_{\ell}t/c$ corresponds to Doppler-induced delay, and under far-field conditions $\tau^{\ell}_{u,\phi}\approx (ud\sin\phi_{\ell})/c$ denotes the spatial delay.

Mixing the received waveform $z_{u}(t)$ with the transmitted chirp $x(t)$ yields the intermediate frequency (IF) signal
\par\noindent\small
\begin{align*}
y_{u}(t)=x(t)z^{*}_{u}(t)=\sum_{\ell=1}^{L}b_{\ell}^{*}&\exp\!\left(j2\pi \kappa\tau^{\ell}_{u}t\right) 
\exp\!\left(-j\pi \kappa(\tau^{\ell}_{u})^{2}\right)\\
&\;\;\times \exp\!\left(j2\pi f_{0}\tau^{\ell}_{u}\right).
\end{align*}
\normalsize
Sampling at rate $f_{s}$ produces the discrete-time signal
\par\noindent\small
\begin{align}
y_{u}[n]=\sum_{\ell=1}^{L} b_{\ell}^{*}\;\underbrace{\exp\!\left(j2\pi\kappa\tau^{\ell}_{u}\tfrac{n}{f_{s}}\right)}_{\text{Term A}}
\underbrace{\exp\!\left(-j\pi\kappa(\tau^{\ell}_{u})^{2}\right)}_{\text{Term B}}
\underbrace{\exp\!\left(j2\pi f_{0}\tau^{\ell}_{u}\right)}_{\text{Term C}},
\label{eq:disc_if}
\end{align}
\normalsize
with $0\leq n \leq N-1$, where $N=\lceil f_{s}T_c\rceil$.

\textbf{Term A:} Define the normalized range frequency $\Psi^{\ell}_{r}\triangleq \kappa\tau^{\ell}_{r}/f_{s}$ and the normalized spatial frequency $\Psi^{\ell}_{\phi}\triangleq d\sin(\phi_{\ell})/\lambda$. Using $\kappa=B/T_c$, where $B=\beta f_{0}$ is the bandwidth fraction, we can expand
\par\noindent\small
\begin{align*}
\exp\!\left(j2\pi\kappa\tau^{\ell}_{u}\tfrac{n}{f_{s}}\right)
=\exp\!\left(j2\pi\Psi^{\ell}_{r}n\right)\exp\!\left(j2\pi\Psi^{\ell}_{\phi}\tfrac{\beta}{N}un\right).
\end{align*}
\normalsize

\textbf{Term B:} The quadratic term is often negligible for typical radar parameters. Substituting \eqref{eq:delay_parts}, we obtain
\par\noindent\small
\begin{align*}
\exp\!\left(-j\pi\kappa(\tau^{\ell}_{u})^{2}\right)
&=\exp\!\left(-j\pi\kappa\tfrac{4r_{\ell}^{2}}{c^{2}}\right)
\exp\!\left(-j\pi\kappa\tfrac{u^{2}d^{2}\sin^{2}\phi_{\ell}}{c^{2}}\right)\\
&\quad\times \exp\!\left(-j2\pi\kappa\tfrac{2r_{\ell}ud\sin\phi_{\ell}}{c^{2}}\right).
\end{align*}
\normalsize
\textcolor{blue}{All of the exponential terms are approximately unity under typical system settings and can therefore be neglected. For instance, consider a scatter located at $r_\ell = 100$ m with $\phi_\ell = 90^{\circ}$, carrier frequency $f_c = 77$ GHz, $U = 128$ virtual array elements, chirp slope parameter $\kappa = 0.1$, propagation speed $c = 3\times10^8$ m/s, and inter-element spacing $d = \lambda/2 \approx 0.0019$ m. Under these conditions, the phase arguments of the quadratic and cross terms are extremely small (i.e., close to zero), resulting in the corresponding exponentials being effectively equal to unity.
} 

\textbf{Term C:} After substituting \eqref{eq:delay_parts},
\par\noindent\small
\begin{align*}
\exp\!\left(j2\pi f_{0}\tau^{\ell}_{u}\right)
=\exp\!\left(j2\pi f_{0}\tau^{\ell}_{r}\right)\exp\!\left(j2\pi \tfrac{ud\sin(\phi_{\ell})}{\lambda}\right).
\end{align*}
\normalsize
The range-dependent factor can also be absorbed into $\tilde{b}_{\ell}$.

\subsubsection*{Equivalent IF Model}
Combining the above, the discrete IF signal in \eqref{eq:disc_if} becomes
\par\noindent\small
\begin{align}
y_{u}[n]\approx \sum_{\ell=1}^{L}\tilde{b}_{\ell}\,
\exp\!\left(j2\pi\Psi^{\ell}_{r}n\right)
&\exp\!\left(j2\pi\Psi^{\ell}_{\phi}u\right)
\underbrace{\exp\!\left(j2\pi\tfrac{\beta}{N}\Psi^{\ell}_{\phi}un\right)}_{\text{SW coupling}}
\nonumber\\&+w_{u}[n], \label{eq:2d_swb}
\end{align}
\normalsize
where $\tilde{b}_{\ell}=b_{\ell}^{*}\exp(j\pi\kappa(\tau^{\ell}_{r})^{2})\exp(-j2\pi f_{0}\tau^{\ell}_{r})$ and $w_{u}[n]\sim\mathcal{CN}(0,\sigma^{2})$.  

The spatial wideband (SW) coupling term links the array index $u$ with the fast-time index $n$. Its impact increases with bandwidth fraction $\beta$, number of array elements $U$, and target angle $\phi_{\ell}$. For small $\beta$, the coupling vanishes and the model reduces to
\par\noindent\small
\begin{align}
y_{u}[n]=\sum_{\ell=1}^{L}\tilde{b}_{\ell}\exp\!\left(j2\pi\Psi^{\ell}_{r}n\right)\exp\!\left(j2\pi\Psi^{\ell}_{\phi}u\right)+w_{u}[n].
\label{eq:2d_nb}
\end{align}
\normalsize

Thus, the radar scene is fully described by the set $\{L, \tilde{b}_{\ell}, \Psi^{\ell}_{r}, \Psi^{\ell}_{\phi}\}$. Traditional 2D spectral methods can recover this signature under the narrowband assumption \eqref{eq:2d_nb}, but fail under the wideband model \eqref{eq:2d_swb}, where range-angle coupling must be explicitly treated.

\section{Proof of Theorem~\ref{thm:SWE_spreading}}

\begin{proof}
\textcolor{blue}{%
We provide a detailed proof based on the Dirichlet-kernel representation of the 2D DFT of the
spatial wideband channel. We focus on a single path $i$ and drop the index $i$ whenever this
does not cause confusion.}

\textcolor{blue}{\textit{Step 1: Single-path model and 2D IDFT.}}
\textcolor{blue}{%
From~\eqref{eq:discrete_space_frequency_channel}, the $(m,n)$th frequency-domain coefficient of path $i$ can be written as}
\begin{equation}
H(m,n)
=
\widetilde{\beta}\,
\e{-j2\pi m\widetilde{\phi}}
\e{-j2\pi n\widetilde{\tau}}
\e{-j2\pi \frac{\alpha}{N}mn\widetilde{\phi}},
\label{eq:single_path_model_SWE}
\end{equation}
\textcolor{blue}{%
where $\widetilde{\phi}\equiv \widetilde{\phi}_i$ and $\widetilde{\tau}\equiv \widetilde{\tau}_i$.
The 2D IDFT (angle-delay) response is}
\begin{equation}
G(k,l)
=
\frac{1}{\sqrt{MN}}
\sum_{m=0}^{M-1}\sum_{n=0}^{N-1}
H(m,n)\,
\e{j2\pi\left(\frac{k m}{M}+\frac{l n}{N}\right)}.
\label{eq:Gkl_def_SWE}
\end{equation}

\textcolor{blue}{%
Substituting~\eqref{eq:single_path_model_SWE} into~\eqref{eq:Gkl_def_SWE} and rearranging the
terms yields}
\begin{align}
G(k,l)
&=
\frac{\widetilde{\beta}}{\sqrt{MN}}
\sum_{m=0}^{M-1}\sum_{n=0}^{N-1}
\e{-j2\pi m\widetilde{\phi}}
\e{-j2\pi n\widetilde{\tau}}\nonumber\\
&\e{-j2\pi \frac{\alpha}{N}mn\widetilde{\phi}}
\e{-j2\pi\left(\frac{k m}{M}+\frac{l n}{N}\right)} \nonumber\\
&=
\frac{\widetilde{\beta}}{\sqrt{MN}}
\sum_{m=0}^{M-1}
\e{-j2\pi m\left(\widetilde{\phi}-\frac{k}{M}\right)}\nonumber\\
&\sum_{n=0}^{N-1}
\e{-j2\pi n\left(\widetilde{\tau}+\frac{\alpha m}{N}\widetilde{\phi}-\frac{l}{N}\right)}.
\label{eq:Gkl_group_n_SWE}
\end{align}

\textcolor{blue}{\textit{Step 2: Dirichlet kernel representation and its properties.}}
\textcolor{blue}{%
Define}
\begin{equation}
y_m(l)
\triangleq
\widetilde{\tau}+\frac{\alpha m}{N}\widetilde{\phi}-\frac{l}{N},
\label{eq:ym_def_SWE}
\end{equation}
\textcolor{blue}{%
and the $N$-point Dirichlet kernel}
\begin{equation}
D_N(x) \triangleq \sum_{n=0}^{N-1}\e{-j2\pi n x}.
\label{eq:DN_def_SWE}
\end{equation}
\textcolor{blue}{%
Then the inner sum in~\eqref{eq:Gkl_group_n_SWE} is $D_N(y_m(l))$, and we can write}
\begin{equation}
G(k,l)
=
\frac{\widetilde{\beta}}{\sqrt{MN}}
\sum_{m=0}^{M-1}
\e{-j2\pi m\left(\widetilde{\phi}-\frac{k}{M}\right)}
D_N\!\big(y_m(l)\big).
\label{eq:Gkl_DN_form_SWE}
\end{equation}

\textcolor{blue}{%
The $Q$-point Dirichlet kernel admits the closed form}
\begin{equation}
D_Q(x)
=
\e{-j\pi(Q-1)x}
\frac{\sin(\pi Q x)}{\sin(\pi x)}.
\label{eq:Dirichlet_explicit_SWE}
\end{equation}
\textcolor{blue}{%
It is well known that:
(i) $|D_Q(x)|$ attains its maximum $Q$ at $x=0$;
(ii) its main lobe is concentrated in a region $|x|\lesssim 1/Q$; and
(iii) for $|x|$ outside this region, the magnitude is uniformly bounded by a constant. 
Formally, there exist constants $c_0>0$ and $C_1,C_2>0$ (independent of $M$ and $N$) such that}
\begin{equation}
|x|\le \frac{c_0}{Q}
\;\Rightarrow\;
|D_Q(x)| \ge C_1 Q,
\qquad
|x|> \frac{c_0}{Q}
\;\Rightarrow\;
|D_Q(x)| \le C_2.
\label{eq:Dirichlet_bounds_SWE}
\end{equation}

\textcolor{blue}{\textit{Step 3: Delay spreading at a fixed coarse angle bin.}}
\textcolor{blue}{%
We next characterize delay spreading at the coarse angle bin $k=\hat{k}_i$, where
$\hat{k}_i/M$ is the nearest DFT grid to $\widetilde{\phi}$. Evaluating~\eqref{eq:Gkl_DN_form_SWE}
at $k=\hat{k}_i$ gives}
\begin{equation}
G(\hat{k}_i,l)
=
\frac{\widetilde{\beta}}{\sqrt{MN}}
\sum_{m=0}^{M-1}
\e{-j2\pi m\left(\widetilde{\phi}-\frac{\hat{k}_i}{M}\right)}
D_N\!\big(y_m(l)\big).
\label{eq:G_khat_l_SWE}
\end{equation}

\textcolor{blue}{%
For the $i$th path to contribute significantly to delay bin $l$, at least one of the terms
$D_N(y_m(l))$ must be in its main lobe, i.e.,}
\begin{equation}
\big|y_m(l)\big| \le \frac{c_0}{N}
\quad\text{for some } m\in\{0,\dots,M-1\}.
\label{eq:main_lobe_delay_condition_SWE}
\end{equation}
\textcolor{blue}{%
Using the definition in~\eqref{eq:ym_def_SWE}, this condition becomes}
\begin{equation}
\left|
\widetilde{\tau}+\frac{\alpha m}{N}\widetilde{\phi}-\frac{l}{N}
\right|
\le \frac{c_0}{N}
\quad\text{for some }m.
\label{eq:main_lobe_delay_condition_expanded_SWE}
\end{equation}

\textcolor{blue}{%
Equivalently, for each fixed $m$ the integer index $l$ must satisfy}
\begin{equation}
\frac{l}{N}
\in
\Big[
\widetilde{\tau}+\tfrac{\alpha m}{N}\widetilde{\phi}-\tfrac{c_0}{N},\;
\widetilde{\tau}+\tfrac{\alpha m}{N}\widetilde{\phi}+\tfrac{c_0}{N}
\Big].
\label{eq:l_over_N_interval_SWE}
\end{equation}
\textcolor{blue}{%
Define}
\begin{equation}
g(m)\triangleq \widetilde{\tau}+\frac{\alpha m}{N}\widetilde{\phi},
\label{eq:g_of_m_SWE}
\end{equation}
\textcolor{blue}{%
so that the interval in~\eqref{eq:l_over_N_interval_SWE} is
$[\,g(m)-\tfrac{c_0}{N},\,g(m)+\tfrac{c_0}{N}\,]$. Because $\alpha\widetilde{\phi}>0$, 
the function $g(m)$ is strictly increasing in $m$, and the union of these intervals over all
$m$ is}
\begin{equation}
\bigcup_{m=0}^{M-1}
\Big[
g(m)-\tfrac{c_0}{N},\,g(m)+\tfrac{c_0}{N}
\Big]
=
\Big[
g(0)-\tfrac{c_0}{N},\;g(M-1)+\tfrac{c_0}{N}
\Big].
\label{eq:union_delay_intervals_SWE}
\end{equation}

\textcolor{blue}{%
Since $g(0)=\widetilde{\tau}$ and 
$g(M-1)=\widetilde{\tau}+\alpha\frac{M-1}{N}\widetilde{\phi}$, this gives}
\begin{equation}
\frac{l}{N}
\in
\Big[
\widetilde{\tau}-\tfrac{c_0}{N},\;
\widetilde{\tau}+\alpha\frac{M-1}{N}\widetilde{\phi}+\tfrac{c_0}{N}
\Big].
\label{eq:l_over_N_final_SWE}
\end{equation}
\textcolor{blue}{%
Multiplying by $N$, the corresponding range of integer delay indices is}
\begin{equation}
l
\in
\Big[
N\widetilde{\tau}-c_0,\;
N\widetilde{\tau}+\alpha(M-1)\widetilde{\phi}+c_0
\Big].
\label{eq:l_interval_final_SWE}
\end{equation}

\textcolor{blue}{%
Let $\mathcal{S}_i$ denote the set of all integers $l$ satisfying~\eqref{eq:l_interval_final_SWE}.
Its cardinality is bounded by}
\begin{equation}
|\mathcal{S}_i|
\le
\alpha (M-1)\widetilde{\phi}+2c_0+1
=
\alpha M\widetilde{\phi}+O(1).
\label{eq:Si_size_bound_SWE}
\end{equation}

\textcolor{blue}{%
By construction, for all $l\notin\mathcal{S}_i$ we have $|y_m(l)|>c_0/N$ for every $m$, so
$|D_N(y_m(l))|\le C_2$ by~\eqref{eq:Dirichlet_bounds_SWE}. Using~\eqref{eq:G_khat_l_SWE}, we get}
\begin{align}
|G(\hat{k}_i,l)|
&\le
\frac{|\widetilde{\beta}|}{\sqrt{MN}}
\sum_{m=0}^{M-1}
|D_N(y_m(l))|
\le
\frac{|\widetilde{\beta}|}{\sqrt{MN}}
\sum_{m=0}^{M-1} C_2 \nonumber\\
&=
C_2|\widetilde{\beta}|\sqrt{\frac{M}{N}},
\qquad l\notin\mathcal{S}_i.
\label{eq:G_outside_Si_bound_SWE}
\end{align}

\textcolor{blue}{%
On the other hand, if $l\in\mathcal{S}_i$, then by definition there exists some $m_0$ such that
$|y_{m_0}(l)|\le c_0/N$. For that index $m_0$, we have from~\eqref{eq:Dirichlet_bounds_SWE} that
$|D_N(y_{m_0}(l))|\ge C_1 N$, and therefore}
\begin{equation}
|G(\hat{k}_i,l)|
\ge
\frac{|\widetilde{\beta}|}{\sqrt{MN}}
\,C_1 N
=
C_1|\widetilde{\beta}|\sqrt{\frac{N}{M}}.
\label{eq:G_inside_Si_lower_SWE}
\end{equation}

\textcolor{blue}{%
For sufficiently large $N$, the lower bound in~\eqref{eq:G_inside_Si_lower_SWE} dominates 
the upper bound in~\eqref{eq:G_outside_Si_bound_SWE}. Hence, for any $0<\varepsilon<1$, there
exists $N_0(\varepsilon)$ such that for all $N\ge N_0(\varepsilon)$ and all 
$l\notin\mathcal{S}_i$,}
\begin{equation}
\frac{|G(\hat{k}_i,l)|}{\max_{l'}|G(\hat{k}_i,l')|} < \varepsilon.
\label{eq:epsilon_threshold_SWE}
\end{equation}
\textcolor{blue}{%
This implies $\mathcal{L}_i(\varepsilon)\subseteq\mathcal{S}_i$ and therefore}
\begin{equation}
|\mathcal{L}_i(\varepsilon)| \le |\mathcal{S}_i|
\le \alpha M|\widetilde{\phi}| + C(\varepsilon),
\label{eq:L_subset_S_bound_SWE}
\end{equation}
\textcolor{blue}{%
for some constant $C(\varepsilon)$ absorbing the $O(1)$ term in~\eqref{eq:Si_size_bound_SWE}.
This proves the delay-domain bound in~\eqref{eq:Li_Ki_bounds_SWE}.}

\textcolor{blue}{\textit{Step 4: Angular spreading via $M$-point Dirichlet kernel.}}
\textcolor{blue}{%
An entirely analogous argument applies in the angle domain if we first group~\eqref{eq:Gkl_def_SWE}
with respect to $n$ and use the $M$-point Dirichlet kernel. This yields a condition of the form}
\begin{equation}
\left|
M\Big(1+\frac{\alpha n}{N}\Big)\widetilde{\phi}-k
\right|
\le c_0
\quad\text{for some }n\in\{0,\dots,N-1\},
\label{eq:main_lobe_angle_condition_SWE}
\end{equation}
\textcolor{blue}{%
which defines an interval of integer angle indices of length $\alpha M\widetilde{\phi}+O(1)$.
Following the same main-lobe versus sidelobe comparison as in Step~3, we obtain}
\begin{equation}
|\mathcal{K}_i(\varepsilon)| \le \alpha M|\widetilde{\phi}| + C(\varepsilon).
\label{eq:Ki_bound_SWE}
\end{equation}
\textcolor{blue}{%
Combining~\eqref{eq:L_subset_S_bound_SWE} and~\eqref{eq:Ki_bound_SWE} establishes the
double inequality~\eqref{eq:Li_Ki_bounds_SWE}.}

\textcolor{blue}{\textit{Step 5: Closed-form spreading limit and asymptotic scaling.}}
\textcolor{blue}{%
The inequalities in~\eqref{eq:Li_Ki_bounds_SWE} show that both the angular and delay support
sizes grow linearly with $\alpha M|\widetilde{\phi}_i|$, up to a constant shift. Thus it is
natural and conservative to define the SWE spreading limit $L_i$ as in
\eqref{eq:Li_closed_form_SWE}. The asymptotic scaling
$|\mathcal{K}_i(\varepsilon)|,|\mathcal{L}_i(\varepsilon)|\sim \alpha M|\widetilde{\phi}_i|$
follows immediately for $M,N\to\infty$ with fixed $\alpha$, which completes the proof.}
\end{proof}

\ifCLASSOPTIONcaptionsoff
\newpage
\fi
\bibliographystyle{IEEEtran}
\bibliography{Biblli1}
\end{document}